\documentstyle[psfig,times]{mn}  

\begin{document}
\def\reef{\par\noindent\hang}
\def\etal{et al.\ }
\def\eg{{\em eg.\ }}
\def\etc{{\em etc.\ }}
\def\ie{{\em i.e.\ }}

\def\spose#1{\hbox to 0pt{#1\hss}}
\def\approxlt{\mathrel{\spose{\lower 3pt\hbox{$\sim$}}
	\raise 2.0pt\hbox{$<$}}}
\def\approxgt{\mathrel{\spose{\lower 3pt\hbox{$\sim$}}
	\raise 2.0pt\hbox{$>$}}}
	
\def\Mdot{\hbox{$\dot M$}}
\def\degmark{$^\circ$}
\def\<{\thinspace}
\def\s{\hbox{\phantom{5}}}	
\def\ss{\s\s}		
\def\sss{\ss\s}		
\def\ssss{\ss\ss}	
\def\lit{\obeyspaces\obeylines}
%
\def\arc{{\rm\thinspace arcsec}}
\def\cm{{\rm\thinspace cm}}
\def\ct{{\rm\thinspace ct}}
\def\erg{{\rm\thinspace erg}}
\def\eV{{\rm\thinspace eV}}
\def\g{{\rm\thinspace g}}
\def\G{{\rm\thinspace G}}
\def\ga{{\rm\thinspace gauss}}
\def\K{{\rm\thinspace K}}
\def\keV{{\rm\thinspace keV}}
\def\m{{\rm\thinspace m}}
\def\km{{\rm\thinspace km}}
\def\kpc{{\rm\thinspace kpc}}
\def\Lsun{\hbox{$\rm\thinspace L_{\odot}$}}
\def\rad{{\rm\thinspace rad}}
\def\MeV{{\rm\thinspace MeV}}
\def\Mpc{{\rm\thinspace Mpc}}
\def\Msun{\hbox{$\rm\thinspace M_{\odot}$}}
\def\pc{{\rm\thinspace pc}}
\def\ph{{\rm\thinspace photons}}
\def\s{{\rm\thinspace s}}
\def\yr{{\rm\thinspace yr}}
\def\sr{{\rm\thinspace sr}}
\def\Hz{{\rm\thinspace Hz}}
\def\GHz{{\rm\thinspace GHz}}
\def\W{{\rm\thinspace W}}
\def\cmps{\hbox{$\cm\s^{-1}\,$}}
\def\ctps{\hbox{$\ct\s^{-1}\,$}}
\def\ctpsparcsecsq{\hbox{$\ct\s^{-1}\arc^{-2}\,$}}
\def\cmsq{\hbox{$\cm^2\,$}}
\def\cmcu{\hbox{$\cm^3\,$}}
\def\pHz{\hbox{$\Hz^{-1}\,$}}
\def\pcmcu{\hbox{$\cm^{-3}\,$}}
\def\ergcmcups{\hbox{$\erg\cm^3\ps\,$}}
\def\ergpcmps{\hbox{$\erg\cm^{-3}\s^{-1}\,$}}
\def\ergpcmsqps{\hbox{$\erg\cm^{-2}\s^{-1}\,$}}
\def\ergpspcmsq{\hbox{$\erg\cm^{-2}\s^{-1}\,$}}
\def\ergpspkpcsq{\hbox{$\erg\s^{-1}\kpc^{-2}\,$}}
\def\ergpspA{\hbox{$\erg\s^{-1}\AA^{-1}\,$}}
\def\ergpspcmsqpA{\hbox{$\erg\s^{-1}\cm^{-2}$\AA$^{-1}\,$}}
\def\ergpcmsqpspsqarcsec{\hbox{$\erg\cm^{-2}\s^{-1}\arc^{-2}\,$}}
\def\ergpcmsqpspapsqarcsec{\hbox{$\erg\cm^{-2}\s^{-1}\AA^{-1},\arc^{-2}\,$}}
\def\ergps{\hbox{$\erg\s^{-1}\,$}}
\def\gpcm{\hbox{$\g\cm^{-3}\,$}}
\def\gpcmps{\hbox{$\g\cm^{-3}\s^{-1}\,$}}
\def\gps{\hbox{$\g\s^{-1}\,$}}
\def\WpHz{\hbox{$\W\Hz^{-1}\,$}}
\def\kmps{\hbox{$\km\s^{-1}\,$}}
\def\ksec{\hbox{$ksec\,$}}
\def\Lsunppc{\hbox{$\Lsun\pc^{-3}\,$}}
\def\Msunpc{\hbox{$\Msun\pc^{-3}\,$}}
\def\Msunpkpc{\hbox{$\Msun\kpc^{-1}\,$}}
\def\Msunppc{\hbox{$\Msun\pc^{-3}\,$}}
\def\Msunppcpyr{\hbox{$\Msun\pc^{-3}\yr^{-1}\,$}}
\def\Msunpyr{\hbox{$\Msun\yr^{-1}\,$}}
\def\pcm{\hbox{$\cm^{-1}\,$}}
\def\pcmsq{\hbox{$\cm^{-2}\,$}}
\def\kpcsq{\hbox{$\kpc^{2}\,$}}
\def\pcmsqpkeVps{\hbox{$\cm^{-2}\keV^{-1}\s^{-1}\,$}}
\def\pmsq{\hbox{$\m^{-2}\,$}}
\def\radpmsq{\hbox{$\rad\m^{-2}\,$}}
\def\pcmcuK{\hbox{$\cm^{-3}\K$}}
\def\phps{\hbox{$\ph\s^{-1}\,$}}
\def\phpcmsqps{\hbox{$\ph\cm^{-2}\s^{-1}\,$}}
\def\pHz{\hbox{$\Hz^{-1}\,$}}
\def\pMpc{\hbox{$\Mpc^{-1}\,$}}
\def\pMpccu{\hbox{$\Mpc^{-3}\,$}}
\def\MsunpMpccu{\hbox{$\Msun\Mpc^{-3}\,$}}
\def\ps{\hbox{$\s^{-1}\,$}}
\def\psqcm{\hbox{$\cm^{-2}\,$}}
\def\psr{\hbox{$\sr^{-1}\,$}}
\def\pyr{\hbox{$\yr^{-1}\,$}}
\def\kmpspMpc{\hbox{$\kmps\Mpc^{-1}$}}
\def\Msunpyrpkpc{\hbox{$\Msunpyr\kpc^{-1}$}}

\title{ Multi-wavelength observations of serendipitous
{\sl Chandra} X-ray sources in the field of A\,2390 }

\author[C.S. Crawford et al ]
{\parbox[]{6.in} {C.S. Crawford $^1$, P. Gandhi $^1$, 
A.C. Fabian $^1$, R.J. Wilman $^{1,2}$, R.M. Johnstone $^1$, 
A.J. Barger $^{3,4,5,6}$ and L.L. Cowie $^{3,6}$  \\
\footnotesize
1. Institute of Astronomy, Madingley Road, Cambridge CB3 0HA \\
2. Leiden Observatory, P.O. Box 9513, 2300 RA Leiden, The Netherlands \\
3. Institute for Astronomy, 2680 Woodlawn Drive, Honolulu HI~96822, USA \\
4. Dept. of Astronomy, University of Wisconsin-Madison, 475~N Charter Street, Madison, WI~53706 USA \\ 
5. Hubble Fellow and Chandra Fellow at Large\\
6. Visiting Astronomer, W.M. Keck Observatory, jointly operated by the California Institute
of Technology and the University of California \\}}

\maketitle
\begin{abstract}
We present optical spectra and near-infrared imaging of a sample of
serendipitous X-ray sources detected in the field of {\sl Chandra}
observations of the A\,2390 cluster of galaxies. The sources have
0.5-7\keV\ fluxes of $0.6-8\times10^{-14}$\ergpcmsqps and lie around
the break in the 2-10\keV\ source counts. They therefore are typical
of sources dominating the X-ray Background in that band. Twelve of the
fifteen targets for which we have optical spectra show emission lines,
most of which have soft X-ray spectra. Including photometric redshifts
and published spectra we have redshifts for seventeen of the sources,
ranging from $z\sim0.2$ up to $z\sim 3$ with a peak between $z=1-2$.
Ten of our sources have hard X-ray spectra indicating a spectral slope
flatter than that of a typical unabsorbed quasar. Two hard sources
that are gravitationally lensed by the foreground cluster are obscured
quasars, with intrinsic 2-10\keV\ luminosities of
$0.2-3\times10^{45}$\ergps and absorbing columns of N$_{\rm
H}>10^{23}$\pcmsq. Both were detected by ISO, showing that the
absorbed flux emerges in the far-infrared.
\end{abstract}


\section{Introduction}
The cosmic X-ray Background (XRB) in the energy range 0.5--7\keV\ has
now been mostly resolved into point sources. About 90 per cent of the
0.5--2\keV\ soft XRB has been resolved with {\sl ROSAT} (Hasinger et
al 1998; Lehmann et al 2001), and now {\sl Chandra} has also resolved
more than 80 per cent of the harder 2--7\keV\ band (Mushotzky et al
2000; Hornschemeier et al 2001; Giacconi et al 2001; Barger et al 2001;
Tozzi et al 2001; Brandt et al 2001). Follow-up work on the deep
fields which gave these {\sl Chandra} results is proceeding but full
details including redshifts are only published for about a half, or
less, of the brighter sources detected. The determination of complete
details of the optically and X-ray faintest sources is likely to take
some time.

The harder 2--7\keV\ X-ray source counts flatten below a flux of about
$10^{-14}\ergpcmsqps$ (Mushotzky et al 2000; Giacconi et al 2001),
which means that much of the information about the origin of the bulk
of the XRB is contained in sources around that flux level. Deep X-ray
observations are not required to detect such sources and several tens
are routinely found as serendipitous sources in {\sl Chandra}
exposures of 10--20~ks on other targets. Indeed, for a given total
exposure time, more such sources are found from say 10 separate 20~ks
exposures than will be found from one deep 200~ks exposure, due to the
flatter faint source counts.

We have begun a programme (Fabian et al 2000; Crawford et al 2001) of
studying the serendipitous sources in our {\sl Chandra} cluster
fields, which have typical exposures of about 20~ks. At this flux
level the optically-identified sources are associated roughly equally
with normal quasars, optically-bright galaxies and optically-faint
galaxies (Mushotzky et al 2000). The latter often appear to be of
early-type (Barger et al 2001). All are plausibly powered by active
galactic nuclei (AGN), but
the harder sources are mostly obscured by significant intrinsic
absorption columns ($\sim 10^{21}-10^{23}\psqcm$). Reddening by dust
associated with this absorbing gas can render the AGN itself
undetectable at optical and near-infrared wavelengths. Obscured AGN
plausibly explain the spectral shape of the XRB (Setti \& Woltjer
1989; Madau, Ghisellini \& Fabian 1994; Comastri et al 1995).

Here we report on the field of the cluster A\,2390 (Fabian et
al 2000; Allen, Ettori \& Fabian 2001; Cowie et al 2001), where we
detect 31 serendipitous sources in two separate {\sl Chandra} observations
totalling about 19~ks. Two have stellar optical spectra and of the
remaining 21 optical identifications we have measured spectroscopic
redshifts for 13 and obtain photometric redshifts from optical and
near-infrared images of a further 4. We test the photometric redshift
approach for such sources against the spectroscopic results where both
are available. The resulting redshift distribution stretches out to
$z\sim 3$ with a peak between $z=1-2$. Ten of our sources have hard
X-ray spectra indicating a spectral slope flatter than that of a
typical unabsorbed quasar. X-ray spectral fitting has been carried out
on the 8 brightest sources. Some X-ray variability is detected between
the two observations.

Strong gravitational lensing has enhanced three sources. These were
also detected by ISO in its deepest exposure, which was of A\,2390
(Altieri et al. 1999). Two of these appear to be genuine obscured
quasars.

\section{Observations and results}

\subsection{ Detection of the X-ray serendipitous sources }

The $z=0.228$ cluster A\,2390 was observed on two occasions with the
{\sl Chandra} X-ray observatory: on 2000 October 08 (with sequence
number 800008 and focal plane temperature of $-$120\degmark C) for
9.83~ks; and on 1999 November 05 (seq. no. 800009 and focal plane 
temperature of $-$110\degmark C) for 9.13~ks. The analysis of the X-ray
emission from A\,2390 itself is published elsewhere (Allen, Ettori \&
Fabian 2001); here we investigate the properties of the serendipitous
point sources identified in the field. For each observation, the
cluster was placed $\sim$1~arcmin from the centre of chip ACIS-S3, and
five of the chips (ACIS-23678) were operating. The roll angle differed
only slightly between the two observations ($-$171\degmark for 800008,
$-$190\degmark for 800009), so there is a large overlap between their 
fields of view.

The 800009 dataset was updated for the correct gainfile (as of 2000
October). The lightcurve of the S3 data showed no evidence for flaring
of the background level during either observation, so we use the data
from the full duration of each exposure. The data were exposure-map
corrected to account for the decrease in effective area off-axis. We
did not attempt to mitigate the charge transfer inefficiency, or
filter on event grades, as we wished to retain the option of
extracting and fitting to an X-ray spectrum of the brighter sources.

We searched the total {\sl Chandra} field of view for serendipitous
sources using the Chandra Interactive Analysis of Observations (CIAO)
WAVDETECT detection algorithm. WAVDETECT was run on the data from each
chip in the 0.5-7\keV\ band, with the data at three binnings: the
original unbinned pixels (each of 0.5 arcsec), and binned by two and
four pixels (ie 1 and 2 arcsec bins respectively). We find this is an
efficient way of detecting the widest range of sources, as the best
precision can be obtained for most sources from the unbinned image,
but the fainter or more diffuse sources (especially in chips where the
point spread function (PSF) is more extended) only appear significant
at higher binnings. We experimented with a full range of wavelet
scales (using the $\sqrt2$ sequence of 1, 1.414, 2.0 \ldots 16.0 pixels) to
pick up different sized sources, and set the significance threshold
for sources at $10^{-5}$ for the 2-pixel binning, and at $10^{-6}$ for
the 1-pixel binning and unbinned data.

We discarded all sources with fewer (non-background subtracted) than
10 counts as too marginal, and any found within 20 arcsec of the edge
of each chip. The spacecraft dither is on this scale, and so there is
a danger that some of the counts for sources too near the chip edge
will be lost during the observation. Each source was examined by eye,
in order to discard any that could be due to the prominent stripes
apparent in the background of some of the chips. These stripes appear
at soft energies, and are most noticeable in chip ACIS-S4, and we only
include one (totally unambiguous) source from this chip (A12). Other
chips marginally affected by background stripes are chips I2, I3 and
S2 in the 800009 observation. We end up with a total of 31
serendipitous sources, and they are listed in Table~1. The first
dataset (800009) was observed when the registration between the {\sl
Chandra} coordinates and the real Sky was incorrect. We have corrected
for this using the 12 sources that were detected in common between the
two separate observations. All the coordinates given for the 800009
dataset in Table~1 have been corrected for this $\sim2$~arcsec offset.

\begin{figure}
\psfig{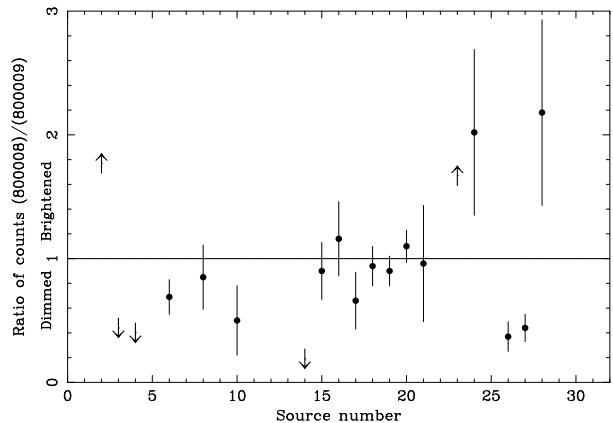}
\caption{\label{fig:dimbright} The ratio of counts in the
serendipitous sources in common between 800009 (earlier) 
and the 800008 (later) observations. One-$\sigma$ errors are shown on
the ratios.  }
\end{figure}

We also estimated the counts in each source in 3 energy bands:
0.5-2\keV\ (soft), 2-7\keV\ (hard) and 0.5-7\keV\ (total). The counts
were taken from a box centred around a source, with a length given by
the square root of the number of pixels in the source cell (as given
from WAVDETECT). The local background was estimated from a concentric
box with a length five times longer. The background box was offset if
it would otherwise spill over the edge of the chip, or would include a
close neighbouring source. Where the source counts (in any band) were
less than twice the error on that value, we replaced the value with an
upper limit set at twice the error. For nearly all the sources the
counts obtained from WAVDETECT and our box statistics agree within the
errors. The only major exceptions are sources A1 -- the most off-axis,
and a very diffuse source -- and A22, which is perhaps rather close to
the edge of the chip. We used the background-subtracted counts in the
hard and soft bands to estimate a soft-to-hard ratio (S/H) for each
source. The higher sensitivity of {\sl Chandra} below 2\keV\ means
that even genuinely hard sources can still show plenty of counts in
our soft band (see eg Crawford \etal 2001). The WAVDETECT results,
counts for each source in the different bands and the S/H ratio are
presented in Table~1. A few of the sources also varied significantly
in brightness over the 11 months between the two observations.
Specifically, sources A2 and A23 brightened over this period, whilst
A3, A4, A6, A14, A26 and A27 all faded (Fig~\ref{fig:dimbright}). The
colours of A6 and A27 also appeared to soften appreciably over this
period.

Our faintest sources typically have 10-11 counts over the full
0.5-7\keV\ energy range; 10 counts corresponds to countrates of
1.017$\times10^{-3}$ \ctps in the 800008 observation and
1.095$\times10^{-3}$ \ctps for the 800009 observation. Assuming a power
law model with slope $\Gamma=1.4$ (similar to the slopes we find when
modelling the spectra of our brighter sources; see section 2.5), that
is absorbed only by the Galactic column ($\sim
6.81\times10^{20}$\pcmsq in this direction; Stark et al 1992) we use
PIMMS to translate source counts into fluxes. A source with a
countrate of $10^{-3}$\ctps observed in a back-illuminated chip (S1 or
S3) is predicted to have a (0.5-7\keV) flux of
6.7$\times10^{-15}$\ergpcmsqps. If observed in a front-illuminated
chip this countrate corresponds to 1.1$\times10^{-14}$\ergpcmsqps.

\subsection{ Infra-red and optical photometry}

We reduced optical images of the field around A\,2390 available from the
archives of the Isaac Newton Group telescopes, and the
Canada-France-Hawaii Telescope. Where the position of a serendipitous
source was not covered in the field of view of one of the archival
datasets, we used the Digitized Sky Survey (DSS; second generation)
images.  The optical data enabled us to preferentially select sources
with no, or a faint, optical identification as most suitable for
imaging in the near-infrared. J, H and K photometry of twelve of the
serendipitous sources from the 800009 observation was obtained on the
the nights of 2000 August 10 and 11 at the United Kingdom Infrared
Telescope (UKIRT), using the UFTI full-array with a 92~arcsec field of
view. A summary of the observations (archival and dedicated) taken are
shown in Table~\ref{tab:obs}. The optical data were flux-calibrated
using standard Landolt stars, except for the CFHT $I$-band
data, which was cross-calibrated from the flux-calibrated WHT $I$-band
image. The standards for the infra-red data were taken from the UKIRT
faint-star catalogue.

Magnitudes of the sources were obtained using PHOT in IRAF, using a
consistent aperture in all the bands in order to provide accurate
colours for later user in photometric redshift fitting. Since the
seeing for the $B$-band was very large, a typical aperture of
4.5~arcsec diameter was chosen for most objects, except for some of
the very faint ones and for objects with close neighbours, for which
an aperture of 2.7~arcsec diameter was chosen, and seeing correction
performed in all the bands. The optical and infra-red magnitudes are
listed in Table~\ref{tab:mags}. Since this is a crowded field, a very
large sky aperture was used estimate the background, relying on PHOT
K-sigma clipping to remove any sources in the background box. A
limiting magnitude was estimated for each source as the 3-$\sigma$ sky
standard deviation multiplied by the square-root of the number of good
pixels in the object aperture. The infra-red images were typically
uncrowded and background estimation could be performed with apertures
that did not typically include other sources.  The magnitudes for our
sources agree well with the magnitudes obtained by Cowie et al. (2001)
when account is taken of the fact that the apertures we use are large
(typically a diameter of 4.5~arcsec to encompass the poor PSF in the
optical) as compared to 2~arcsec diameter used in Cowie et al.

\subsection{Optical spectroscopy}
We took optical spectra of fifteen of the serendipitous background
sources listed in Table~1; the selection of which sources to observe
was fairly random, with a preference only for those with definite
optical identifications, and which were least off-axis in the {\sl
Chandra} observations. The spectra were taken on 2000 September 28 and
29 with the Keck Echelette Spectrograph and Imager (ESI; Epps \&
Miller 1998) used in low resolution mode with a 1~arcsec slit. The
nights were photometric with seeing of about 0.7~arcsec. Targets were
observed at three positions along the slit with 600 sec exposure times
at each position and the median of the observations used to form the
sky. The observations were then sky subtracted, aligned, and the
spectrum extracted. The wavelength calibration was determined from the
night sky emission lines. The final spectral resolution is $17\AA$ and the
data cover the $5000-10000\AA$ wavelength range. Redshifts were then
measured by hand for each object using a standard set of emission and
absorption lines for galaxies and AGN.

\begin{figure}
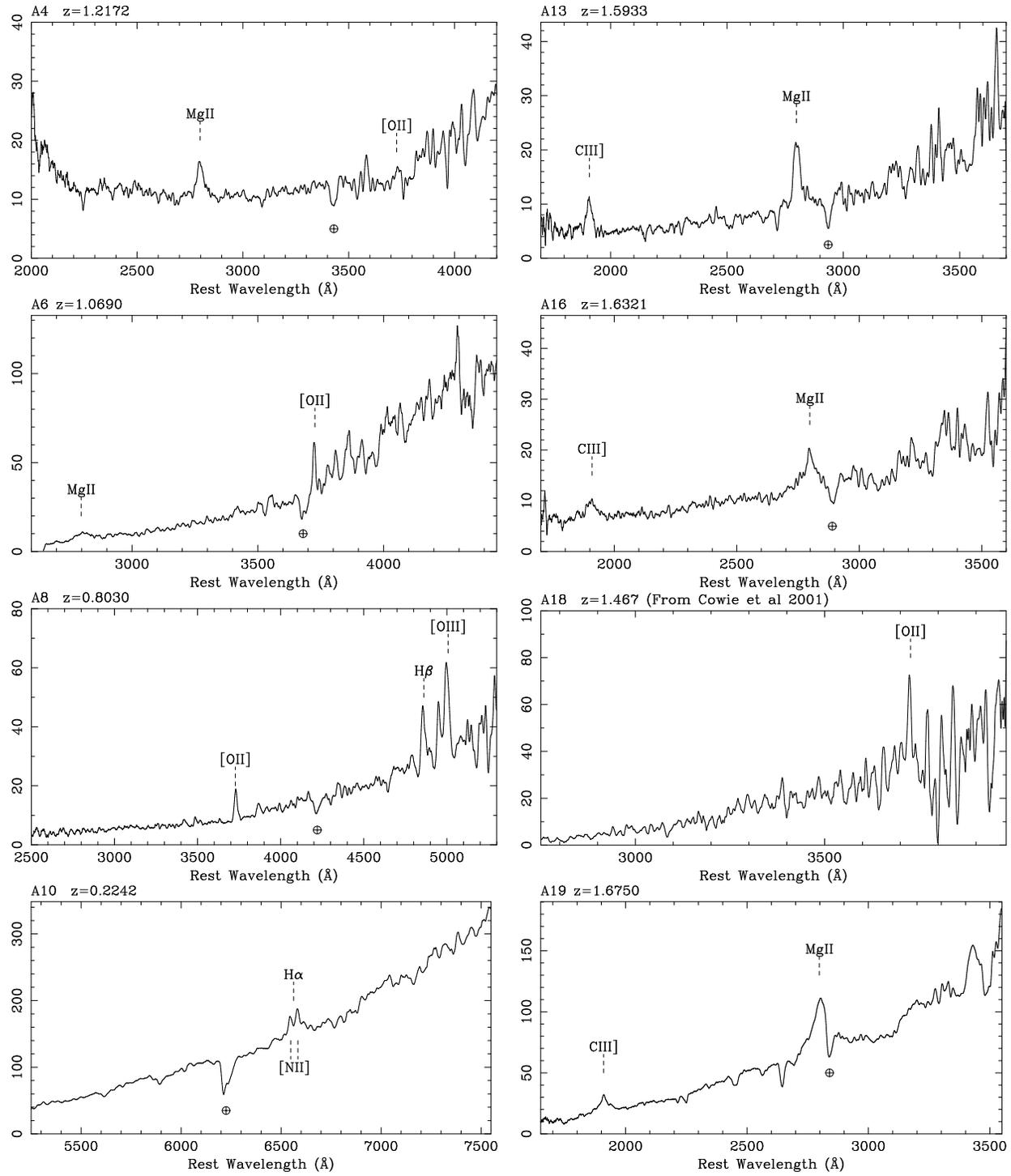

\vbox{
\hbox{
\psfig{figure=a4.ps,width=0.45\textwidth,angle=270}
\hspace{0.25cm}
\psfig{figure=a13.ps,width=0.45\textwidth,angle=270}
}\hbox{
\psfig{figure=a6.ps,width=0.45\textwidth,angle=270}
\hspace{0.25cm}
\psfig{figure=a16.ps,width=0.45\textwidth,angle=270}
}\hbox{
\psfig{figure=a8.ps,width=0.45\textwidth,angle=270}
\hspace{0.25cm}
\psfig{figure=a18.ps,width=0.45\textwidth,angle=270}
}\hbox{
\psfig{figure=a10.ps,width=0.45\textwidth,angle=270}
\hspace{0.25cm}
\psfig{figure=a19.ps,width=0.45\textwidth,angle=270}
}}
\caption{ \label{fig:emsp}
The Keck spectra of the serendipitous sources in the field of A\,2390.
The spectra are not flux-calibrated, and have been smoothed by 20\AA.
The $\oplus$ symbol in each spectrum marks the prominent atmospheric absorption feature at
around 7600\AA. }
\end{figure}

Eleven sources show clear emission-line spectra (Fig~\ref{fig:emsp},
with redshifts ranging from 0.214 to 1.675 (Table~\ref{tab:zphot}). At
least one of these (A20) can be clearly identified as a narrow-line
(ie Seyfert-II-like) AGN. A further two sources (A5 and A25) show a
stellar spectrum, and are not considered in this paper further. We did
not obtain any clear redshift identification for the two remaining
sources, A15 (Fig~\ref{fig:nozsp}) and A18. We do, however, detect the
[OII] emission line in the spectrum of A18 at the
spectroscopically-confirmed redshift of $z=1.467$ from Cowie et al
(2001).  The equivalent widths of the principal emission lines are
tabulated in Table~\ref{tab:ew}, as well as the velocity width of the
MgII$\lambda2798$ emission line.  In most cases the MgII line is
comparatively narrow, although we note that there is evidence for a
broad component in both A16 and A19.

\onecolumn

\addtocounter{figure}{-1}
\begin{figure}
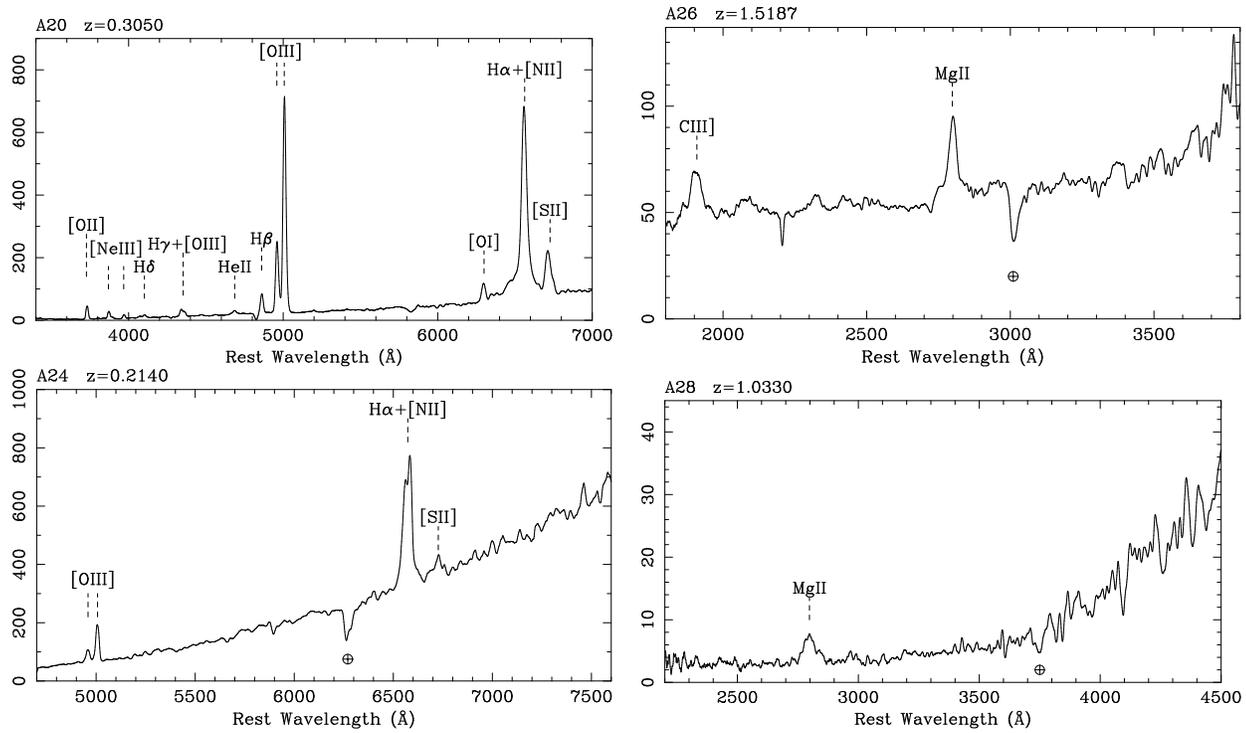

\vbox{
\hbox{
\psfig{figure=a20.ps,width=0.45\textwidth,angle=270}
\hspace{0.25cm}
\psfig{figure=a26.ps,width=0.45\textwidth,angle=270}
}\hbox{
\psfig{figure=a24.ps,width=0.45\textwidth,angle=270}
\hspace{0.25cm}
\psfig{figure=a28.ps,width=0.45\textwidth,angle=270}
}
}
\caption{The Keck spectra of the serendipitous sources in the field of
A\,2390 (ctd). }
\end{figure}

\begin{figure}
\psfig{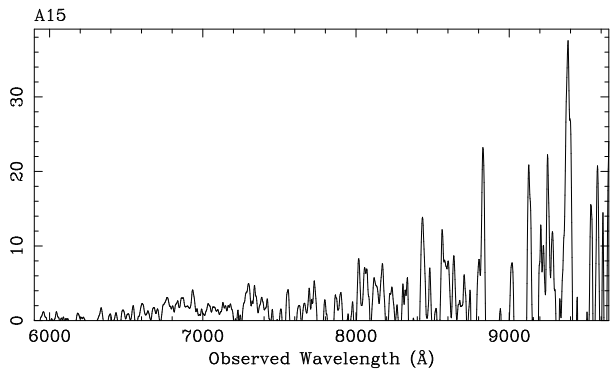}
\caption{\label{fig:nozsp} The Keck spectrum of the serendipitous source A15.}
\end{figure}

\twocolumn

\begin{figure}
\psfig{figure=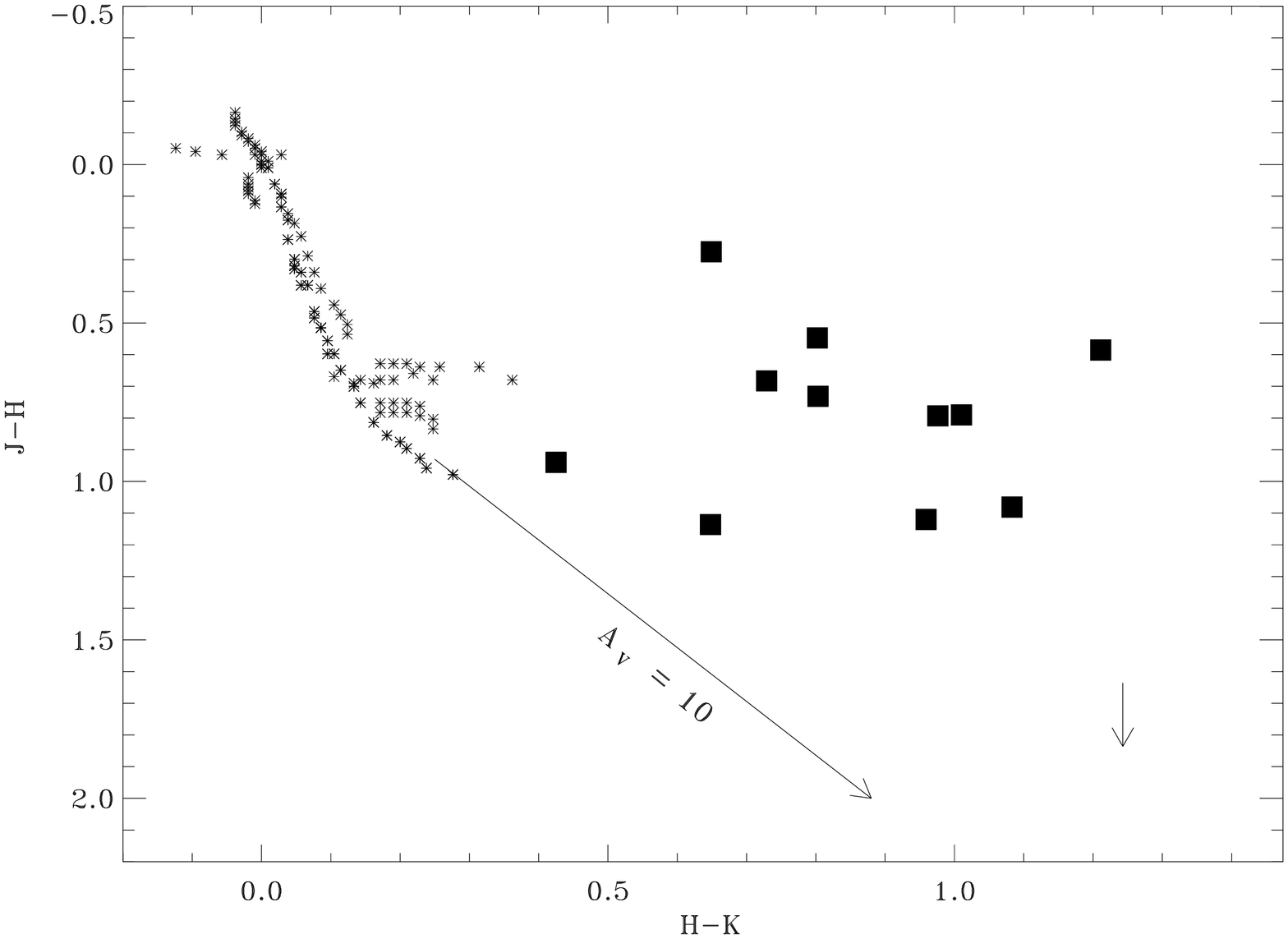,width=0.5\textwidth,angle=90}
\caption{\label{fig:jhk}
The near-infrared colour-colour plot for the A\,2390 background sources
(filled squares). The star symbols show the infrared colours of main
sequence, giant and supergiant stars of all spectral types. The upper
limit is that of A15, in which the large {\sl J-H} break places the
object at z$_{\rm{phot}}=2.78$. }
\end{figure}

\subsection{Photometric Redshifts}

As it is clear from the near-infrared colours (Fig~\ref{fig:jhk}) that
we are not observing stellar objects, we use HYPERZ, a publicly
available code (Bolzonella, Miralles \& Pello 2000) to estimate
redshifts for the sources for which we have near-infrared magnitudes.
HYPERZ matches template spectra to the observed spectral energy
distribution (SED), which is input as the observed magnitudes and
their errors in as many bands as possible. The template spectra are
convolved with the filter response in each of the input bands. The
filter response includes the quantum efficiency of the CCD --
particularly important in the $I$-band where the response of the CCD
falls rapidly toward longer wavelengths around 9000\AA. The infra-red
filter responses were also folded with the atmospheric transmission at
Mauna Kea, which is affected strongly by water vapour absorption features.

\begin{figure}
\psfig{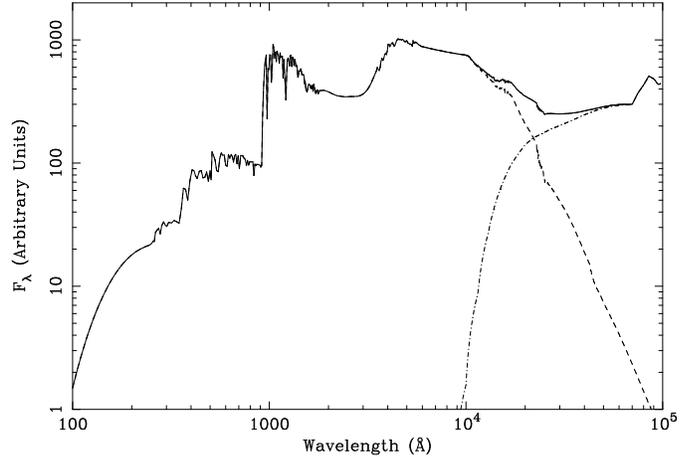}
\caption{\label{fig:cwwscd} The combined spectrum
(solid line) of the CWW Scd template (dashed) with an AGN component
(dash-dot) reddened by DUSTY and normalised to have
50\% of the total flux at 2.2$\mu$m). }
\end{figure}

The redshift, age and internal reddening of the templates are varied
in order to obtain the maximum likelihood $\chi^2$ solution. The
template spectra chosen comprised Bruzual \& Charlot (1993) synthetic
spectra -- with a Solar metallicity and a Miller-Scalo initial mass
function -- and empirical SEDs observed by Coleman, Wu \& Weedman (1980), hereafter referred to as CWW. Both
sets of templates range from a single burst of star-formation (SF)
through ellipticals and spirals (with exponentially decaying SF rates)
to a model for continuous SF in an irregular galaxy. The classification of the CWW spectra is based on the observed morphology of the objects, while the Bruzual \& Charlot model galaxy type is set according to the star-formation timescale.  
In addition, since standard models for the synthesis of the hard X-ray
background (Setti \& Woltjer 1989; Wilman \& Fabian 1999; Wilman,
Fabian \& Nulsen 2000) predict some of our sources to be heavily
obscured AGN, we also created a set of template galaxies with obscured
AGN components. A standard broken power-law AGN continuum (Granato,
Danese \& Franceschini 1997) was assumed over the
near-infrared/optical range and reddened with dust using the radiation
transfer code DUSTY (Ivezi\'{c}, Nenkova \& Elitzur 1999). Since nothing
is known about the dust around these sources without any (a priori)
redshift information (except that the dust is probably hot to hundreds
of degrees Kelvin; see Wilman, Fabian \& Gandhi 2000; hereafter WFG), a
number of models were constructed by varying the dust temperature,
composition, optical depth and normalization of the AGN-SED to
host-galaxy-SED. The reddened AGN component was added to each of the
CWW spectra; one example of this is shown in Fig~\ref{fig:cwwscd},
for the Scd CWW template. With the typical dust properties and
composition found in WFG, all the optical light is assumed to be
depleted and to emerge preferentially in the mid-infrared regime, with some
contribution in the near-infrared. Therefore, the major contribution
of the reddened AGN is to the total $K$-band light.


HYPERZ has a number of options to account for undetected objects, and
we adopted the recommended procedure for medium-deep surveys such as
ours. We set the flux of a non-detected object and its 1-$\sigma$
error equal to F$_{\rm{lim}}/2$ in a filter, where F$_{\rm{lim}}$ is
the flux corresponding to the limiting magnitude in that band.

\begin{figure}
\psfig{figure=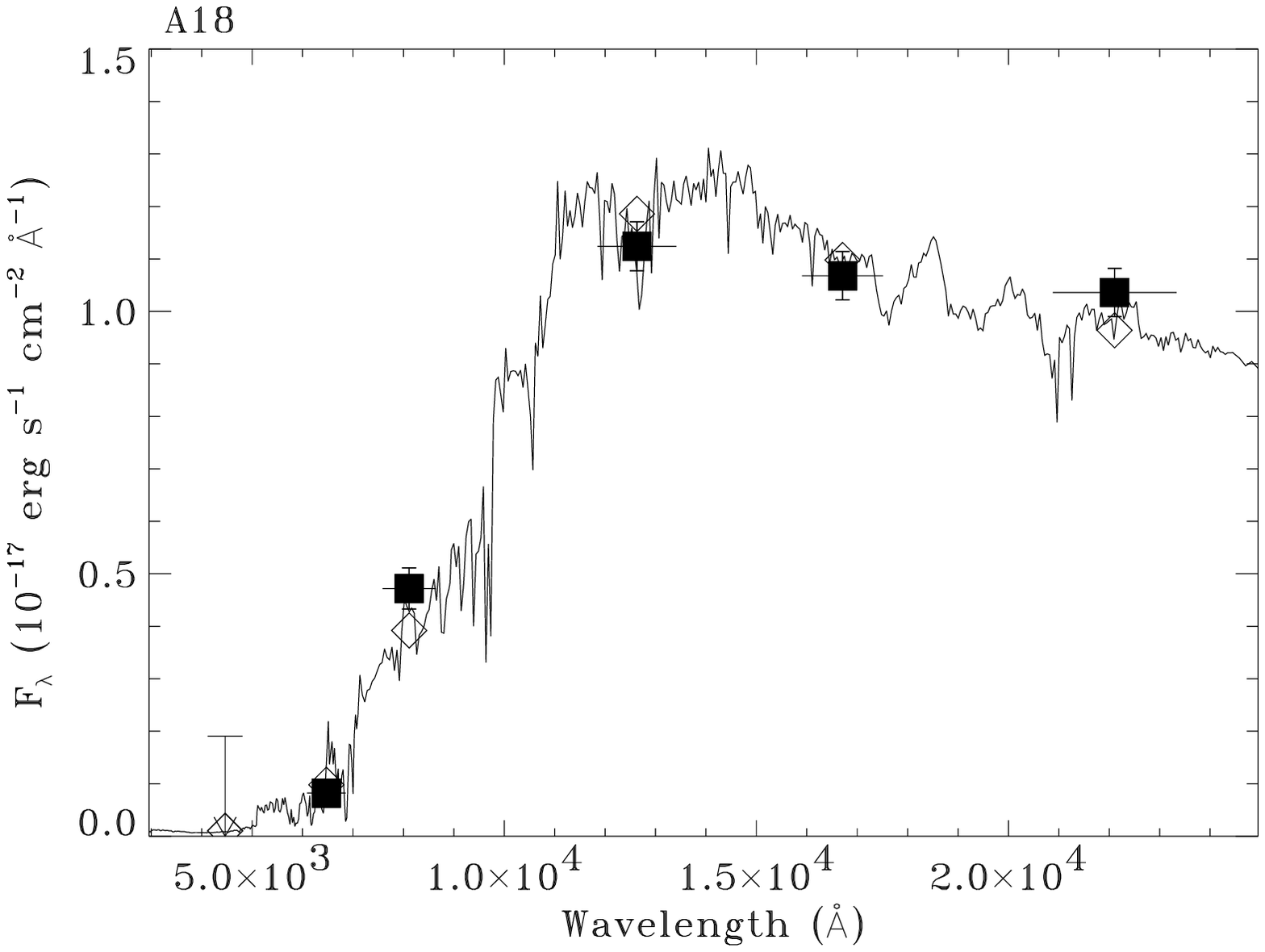,width=0.5\textwidth,angle=0}
\caption{\label{fig:sedfit} The best-fit spectrum (solid line) for A18
compared to the optical and near-infrared fluxes (solid square
markers and {\sl B}-band limit). The galaxy SED is a single Bruzual \& Charlot stellar burst
model at $z_{\rm{phot}}=1.45$, close to $z_{\rm{spec}}=1.467$. The
$x$-errorbars are the bandwidths of the filters calculated by HYPERZ
according to a Gaussian approximation, and the open diamond markers
show the integrated fluxes of the template SED through the individual
filters. } \end{figure}

\begin{figure}
\psfig{figure=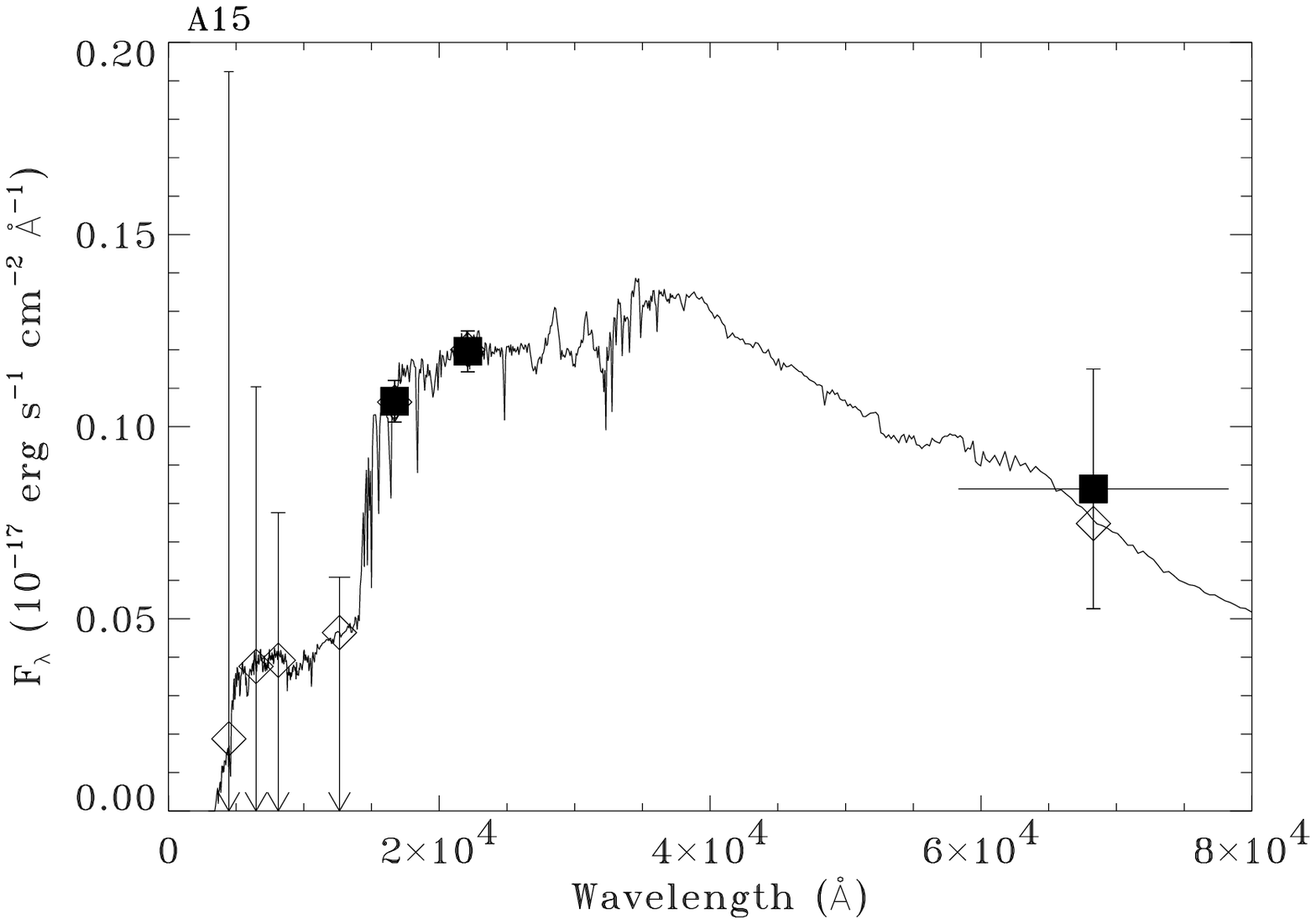,width=0.5\textwidth,angle=0}
\psfig{figure=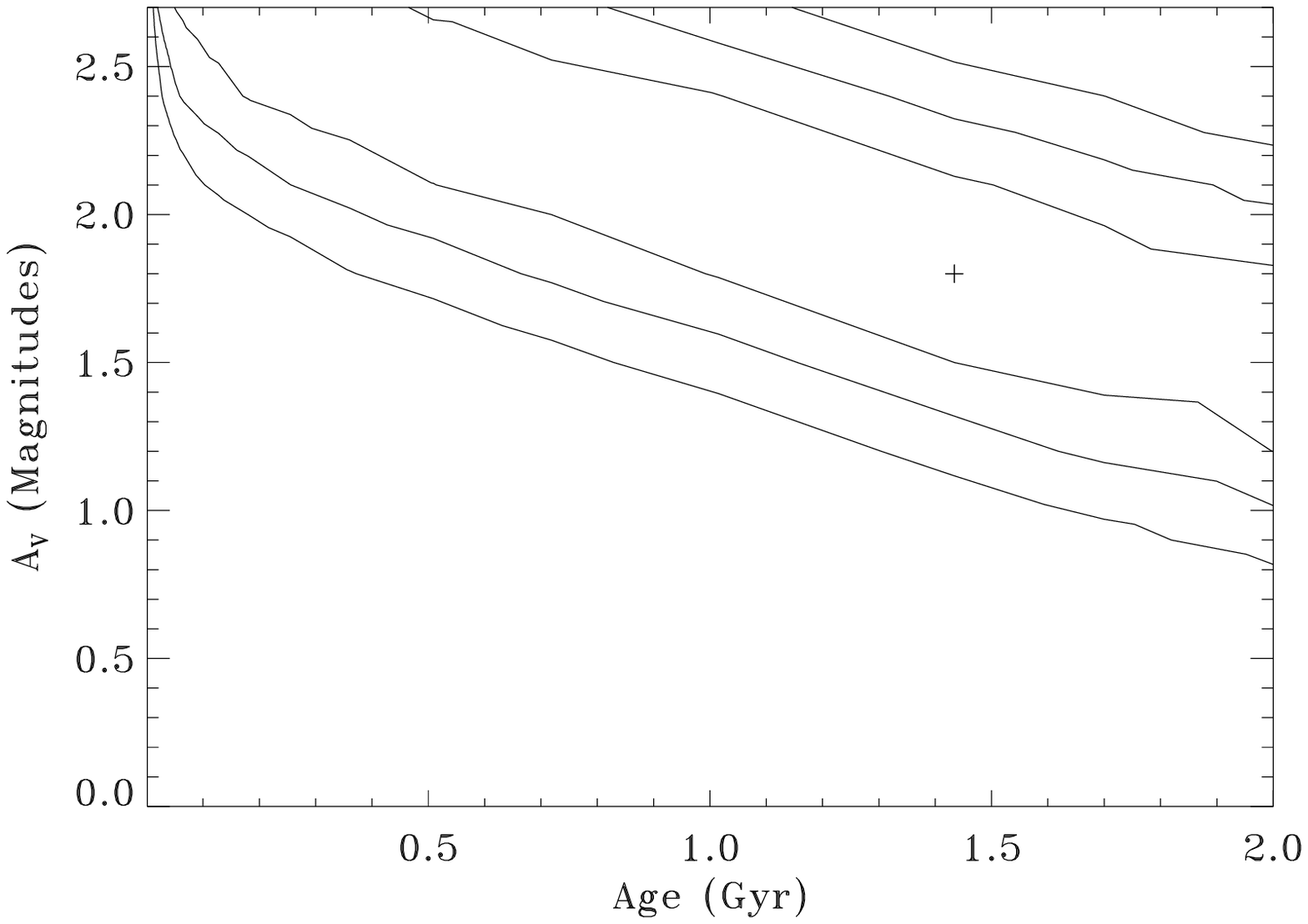,width=0.5\textwidth,angle=0}
\caption{\label{fig:hyperzconf} (Top) The most-likely
HYPERZ SED for A15 compared to the data, based on limits in {\sl B, R,
I} and {\sl J} bands and detections in {\sl H, K} and at 6.7-${\rm{\mu}}$m. The
diamonds show the integrated fluxes of the template SED through the
individual filters.  \newline\noindent (Bottom) Contours of confidence
between the age of the elliptical host galaxy model at $z=2.78$ and
the amount of intrinsic reddening required. The cross indicates the
best model, while the confidence intervals are at probabilities of
68\%, 95\% and 99\%. This illustrates the need for reddening in this
object, for all host galaxy ages up to 2-Gyr.}
\end{figure}

The Galactic column density of $6.81\times10^{20}$ cm$^{-2}$ in the
direction of this cluster corresponds to a reddening $E(B-V)=0.12$.
HYPERZ de-reddens the observed fluxes according to the Galactic
reddening law of Allen (1976). Any reddening internal to the galaxy
itself was assumed to follow the Calzetti reddening law (Calzetti et
al 2000), with a fitting range of $A_V$ between 0.0 and 3.0, in steps
of 0.3. The redshift range considered for most objects was 0.0 to 6.0
in steps of 0.05.  A maximum absolute Vega magnitude of $-$25.0 in the
$B$-band (Bessell filter) for a $H_0=50$; $q_0=0.5$ cosmology was
assumed. The results of the photometric-redshift-fitting procedure are
shown in Table~\ref{tab:zphot}, with examples of the best-fit SED for
A18 shown in Fig~\ref{fig:sedfit} and for A15 in
Fig~\ref{fig:hyperzconf}.  Fig~\ref{fig:hyperzconf} also shows
the contours of confidence for variation of the age of the template
galaxy and the amount of reddening required, demonstrating the need
for reddening for a range of feasible host galaxy ages in this source.

\begin{figure}
\psfig{figure=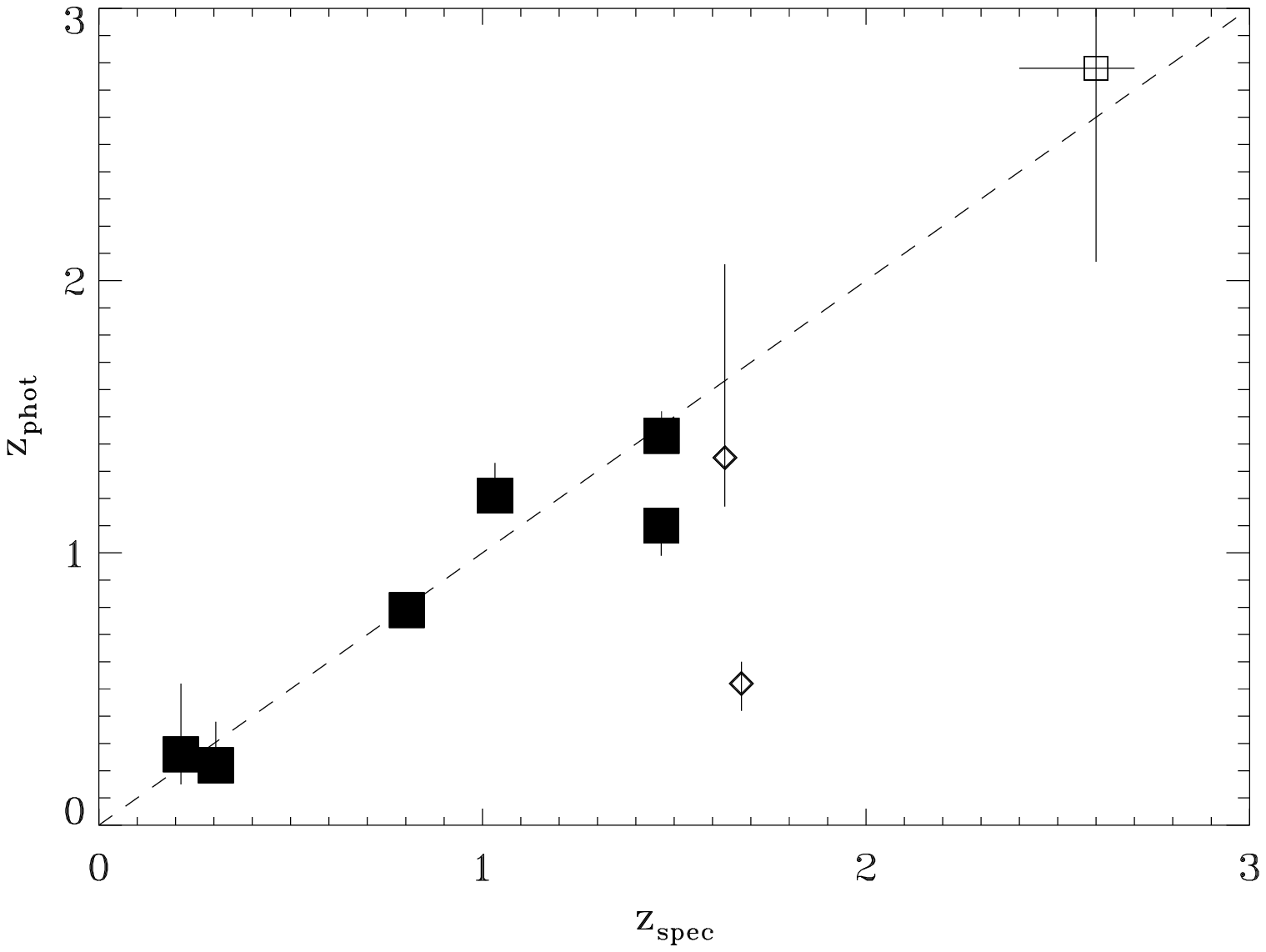,width=0.5\textwidth,angle=0}
\caption{\label{fig:zcompare} Comparison of spectroscopic and
photometric redshifts for nine sources. The $y$-errorbars are taken
from the 90\% confidence limits given by HYPERZ. For the object A15 at
${\rm{z_{phot}=2.78}}$, the value of $z_{\rm spec}=2.6$ quoted for
comparison is the photometric-redshift obtained by Cowie et al (2001).
The two objects marked by unfilled diamonds (A16 and A19) show
other indications of having their light dominated by relatively unobscured quasars, and are
discussed further in the text. }
\end{figure}

The photometric redshift estimate agrees well with the spectroscopic
measurement (from this paper and from Cowie et al 2001: see
Table~\ref{tab:zphot} and Fig~\ref{fig:zcompare}) for five of our
sources: A8, A18, A20, A24 and A28.  In addition, our redshift
estimate for A15 agrees well with the estimate from Cowie et al
(2001), and the spectroscopic redshift for A16 is encompassed within
the 90\% confidence interval of the photometric-redshift fit. We
emphasize that these results were not manipulated to match the
spectroscopic redshifts, but constructed as a \lq blind' test of the
photometric redshift fitting procedure. We have a further three
sources (A12, A14 and A27) for which we obtain a photometric redshift
without spectroscopic confirmation.

A12 was observed during partial cirrus coverage, which leads to a
systematic photometric uncertainty of about 10 per cent in flux when
calibrated against the standard star observed nearest in
time, as opposed to calibration against all stars observed through the
night. Though this in itself is not a significant uncertainty, it
creates a degeneracy in the photometric redshift estimate for this
source, with the primary solution $z_{\rm{phot}}=2.28$ and the
secondary solution $z_{\rm{phot}}=0.66$ being almost equally likely,
depending on the calibration. This is because there is no {\sl I}-band
magnitude available, and we only have relatively-shallow DSS limits in
the {\sl B} and the {\sl R} bands, thus reducing the contraints on
this object. {\sl I}-band photometry would be a strong constraint to
resolve the issue. Wherever quoted in this paper, the primary solution
stated in Table~\ref{tab:zphot} has been assumed.

\subsubsection{A15 at $z\sim2.8$}
\begin{figure}
\psfig{figure=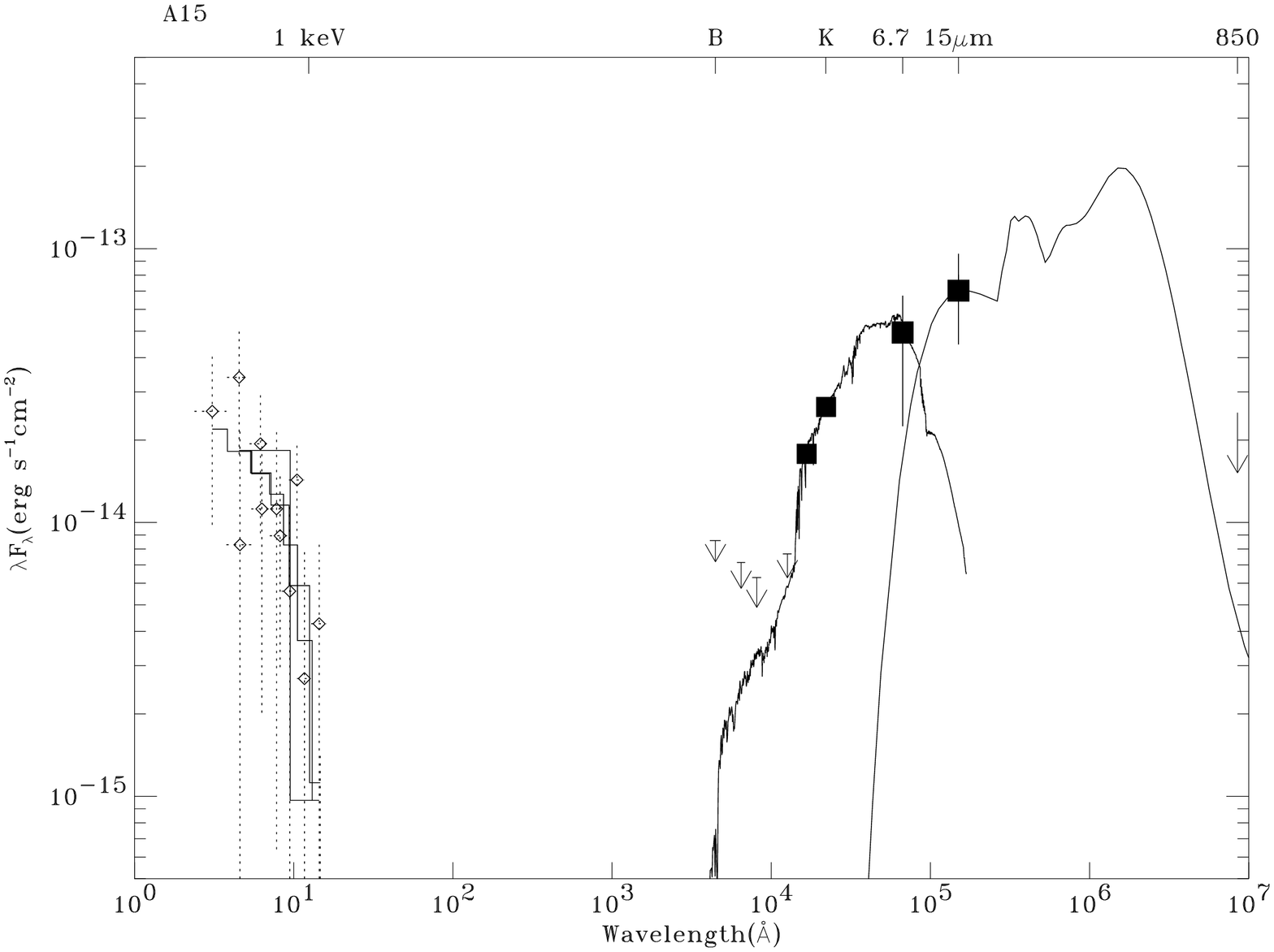,width=0.5\textwidth,angle=90}
\caption{\label{fig:totalsed} The total spectral energy distribution
for A15. The X-ray data are marked by open diamond markers with 
error bars. Our optical and near-infrared values and limits are marked
as solid squares. The best-fit model to the X-ray spectrum (see
Table~\ref{tab:xsp}) is shown as a solid line, and the other solid
lines show the fits obtained in HYPERZ (optical to near-infrared) and
DUSTY (normalized to the 15-${\rm{\mu}}$m flux; see 
text for details). The {\sl SCUBA} upper-limit is shown as the arrow at 850-${\rm{\mu}}$m.
}
\end{figure}

The high redshift of $z\sim2.8$ (with a range of $2.1<z<3.3$)
predicted for A15 is not surprising, given its large $J-H$ colour
(Fig~\ref{fig:jhk}). For most of this redshift range ($z>2.4$) we note
that the only major spectral feature we could expect to fall into our
observed optical waveband would be the CIII] line, perhaps partway
explaining our lack of spectroscopic redshift for this object. This
object lies closest to the central cD galaxy and is strongly lensed by
a factor of 7.8 (Cowie et al. 2001). This is also detected in a
very-deep {\sl ISOCAM} exposure (Altieri et al. 1999) and the fluxes
are reported in L\'{e}monon et al. (1998). WFG used this detection, in
conjunction with a {\sl SCUBA} upper-limit (Fabian et al. 2000) to
model the mid-infrared part of the spectrum as due to radiation
absorbed in the optical-ultraviolet regime and re-emitted at longer
wavelengths. This was done using DUSTY (Ivezi\'{c}, Nenkova \& Elitzur
1999) and is shown as the solid line in the mid- to far-infrared
region of Fig~\ref{fig:totalsed}. Given a central source emitting
power-law radiation (for example), and the density, composition,
amount and optical depth of surrounding dust, DUSTY calculates a
scaled radiative-transfer solution for different dust
temperatures. This spectrum in Fig~\ref{fig:totalsed} has been
redshifted to $z\sim2.8$ (as obtained by HYPERZ for the
optical-to-near-infrared part of the spectrum) and has been normalised
to the 15-$\mu$m flux. Both the
near-infrared and the mid-infrared parts of the total SED are
consistent with the 6.7-$\mu$m flux. This model has a dust-temperature
of 1500 K at the inner-dust radius, and an optical-depth of 40 at
0.3-$\mu$m. Though lower temperature models were formally acceptable,
these underestimate the $SCUBA$ limit by only a factor of a few. Given
that most of the {\sl Chandra} sources are undetected at
sub-millimeter wavelengths (e.g. Fabian et al. 2000), the
850-${\rm{\mu}m}$ flux is likely to lie well below the {\sl SCUBA}
limit. Reducing the temperature significantly begins to cut into the
850-$\mu$m flux and also miss the 6.7-$\mu$m flux completely. (See WFG
for more details. Note that although the redshift presented therein
has since changed, this only makes the conclusion of hot dust even
more robust.) A18 was also detected by ISOCAM (Altieri et al 1999). We
do not produce an SED here since no mid-infrared fluxes have been
published.

\subsubsection{Problem sources}
Our photometric redshift estimates do not agree so well with the
spectroscopic redshifts of A16 and A17, and the estimate is
particularly bad for A19; in all three cases the secondary solution to
the photometric redshift is no better. A16 is the closest match of
these three sources, with the 90\% confidence interval of the
photometric-redshift-fitting encompassing the spectroscopic
measurement.

\begin{figure}
\fbox{\psfig{figure=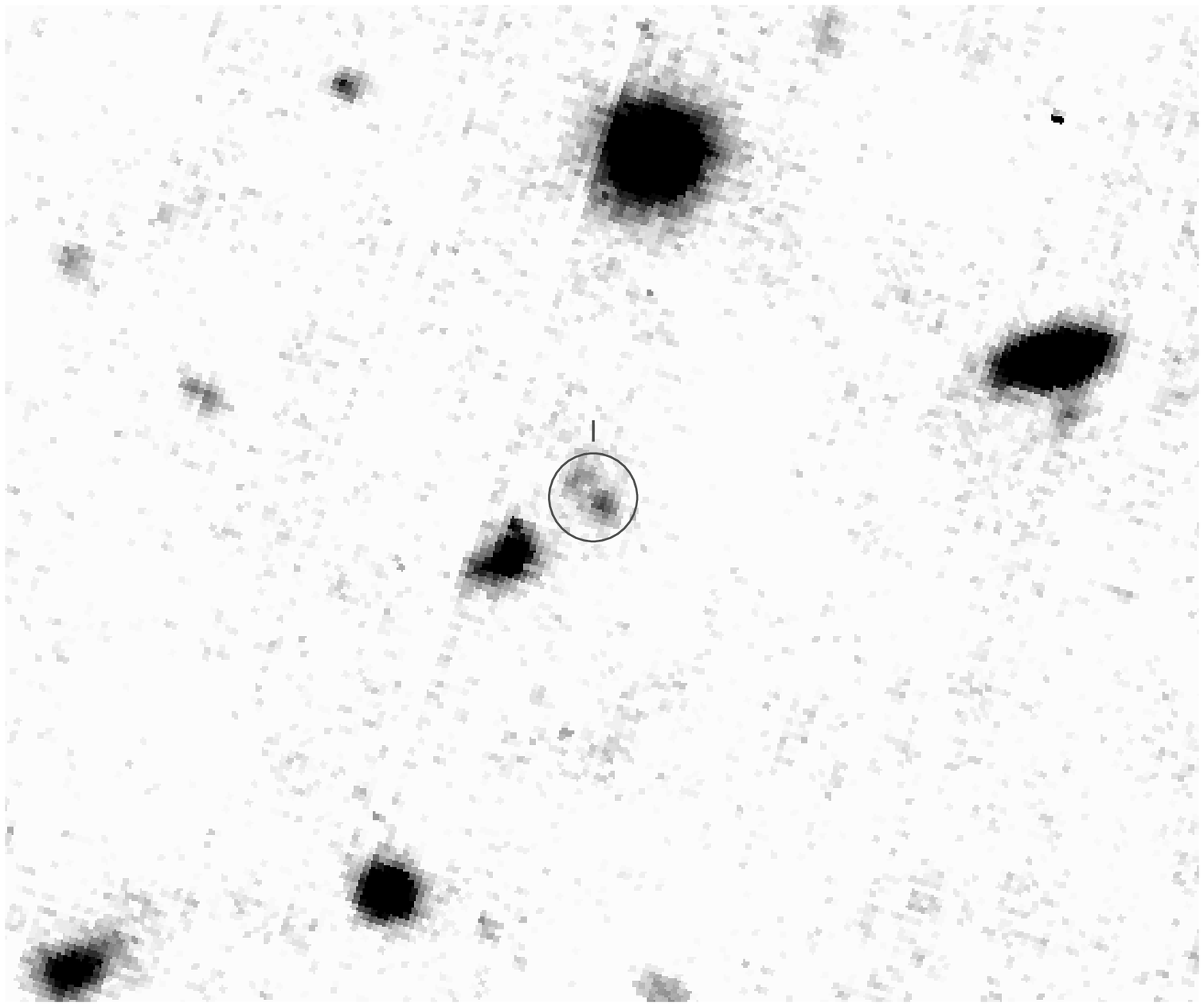,width=0.4\textwidth,angle=0}}
\caption{\label{fig:a17im} {\sl CFHT I}-band image of A17 showing a double structure. In addition
there is confusion with the neighbour to the south-east in the {\sl B}
and the {\sl R}-band images, where the seeing is much worse. The image
side is 30-arcsec long and North is to the top, East to the left. }
\end{figure}

Part of the redshift mismatch for A17 is due to its proximity to a
bright neighbour about 3 arcsec to the south-east (Fig~\ref{fig:a17im}). A small aperture
(2.7~arcsec diameter) was used to
extract the flux, but the source is not well resolved from its
neighbour in the {\sl B}- and {\sl R}-band data due to the very
poor seeing at the time of the observations. Thus the {\sl B}- and
{\sl R}-band magnitudes may be inaccurate due to
contamination from the neighbour and loss of source flux outside the small aperture. The source and its neighbour are
well resolved in both the UKIRT and {\sl CFHT} observations; in
addition, the {\sl I}- and {\sl H}-band images show the source itself
to have a second component $\sim1$~arcsec to the NE
(Fig~\ref{fig:a17im}). This second component is probably too faint to
be a serious source of contamination to the observed magnitudes. 
We speculate that the small NE component could be an extended
($\sim8$\kpc\ from the source) cloud of emission-line gas associated
with the source. If it were at the same redshift as the source of
$z_{\rm phot}$, then strong line emission from [OII] and [NII] and/or
H$\alpha$  could
account for this feature showing up only in the {\sl I}- and {\sl
H}-bands. We note, though, that the {\sl on-source} spectrum of Cowie et al
(2001) -- which may well not include this NE component -- does not
show strong [OII], and any [OIII] line emission from this would
emerge in the {\sl J}-band. 

\begin{figure}
\psfig{figure=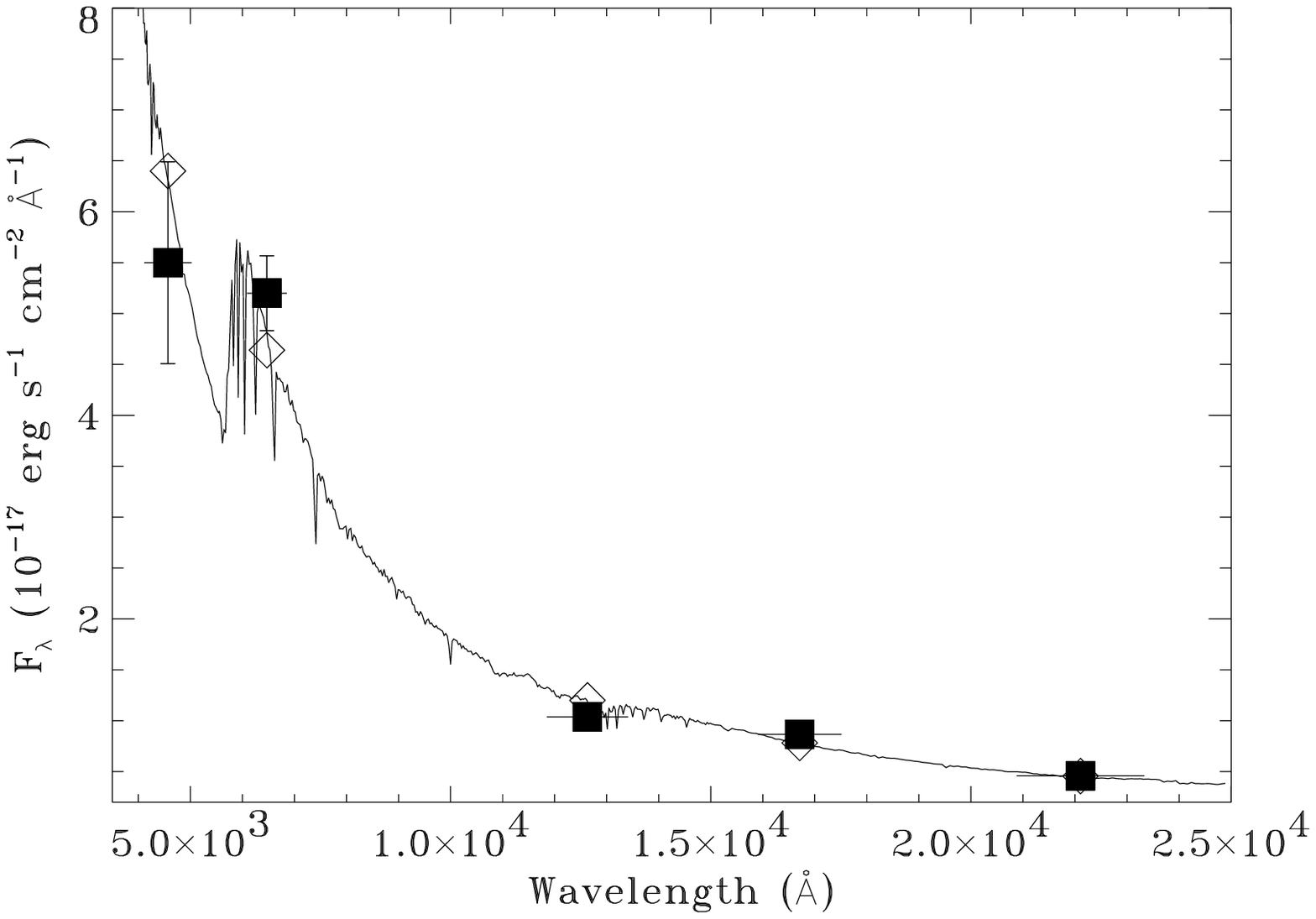,width=0.4\textwidth,angle=0}
\psfig{figure=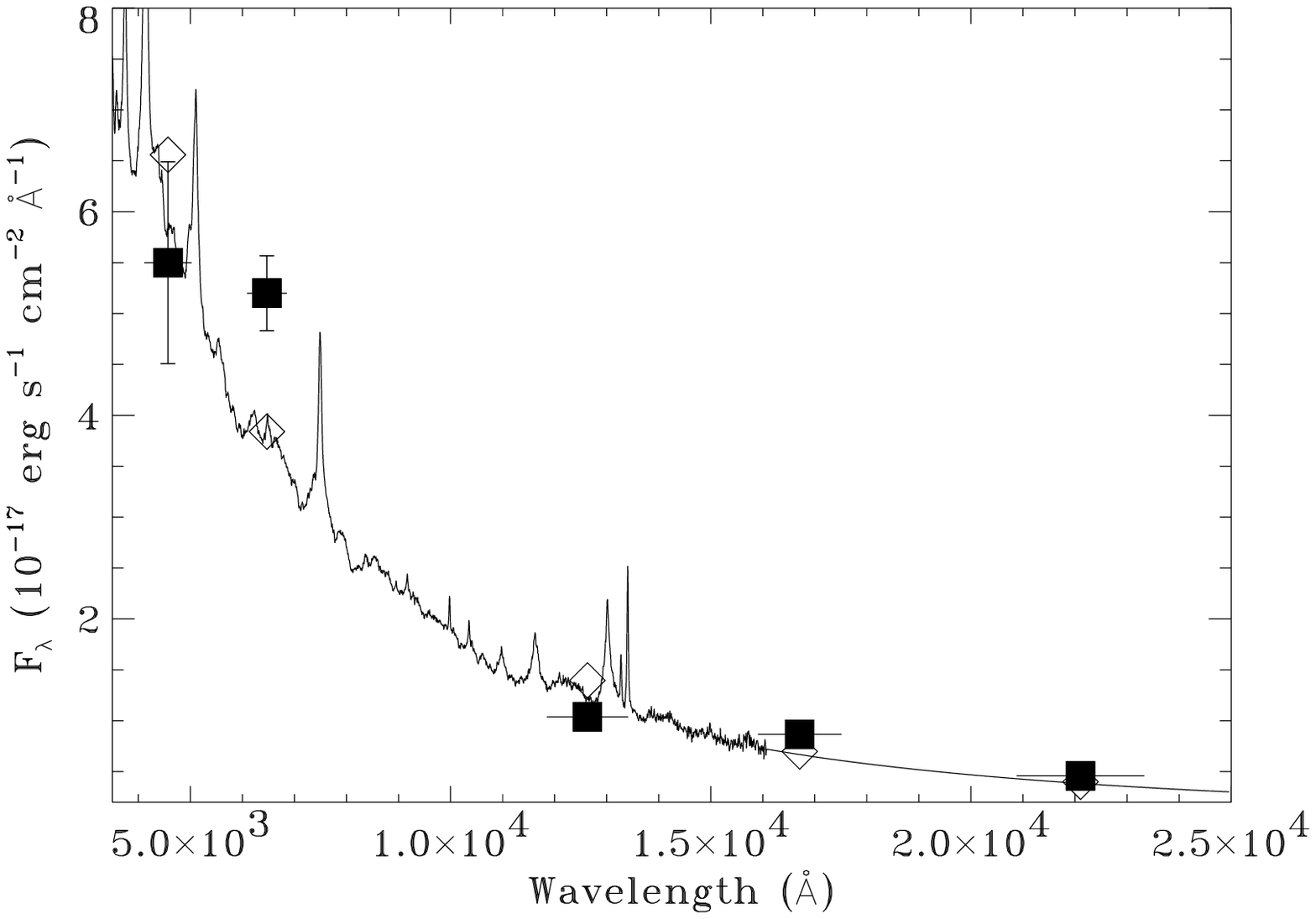,width=0.4\textwidth,angle=0}
\caption{\label{fig:a19sed} The
best-fit SED to A19 using all templates gives a young (260-Myr-old)
system at ${\rm{z_{phot}}}=0.52$ (top), while an empirical QSO
template fixed at ${\rm{z=z_{spec}}}$ and a small $A_V=0.2$ has a
similar slope to the observed fluxes, with some deviation in the {\sl
R}-band.\noindent \newline}
\end{figure}

A19 appears point-like in the near-infrared images, with a
full-width-half-maximum close to the seeing values, and is easily
resolved from any neighbouring sources. Its brightness at optical
wavelengths implies that there are no obvious breaks in the broadband
spectrum (see Fig~\ref{fig:a19sed}) to anchor the photometric redshift
fitting. Moreover, the lack of an archival {\sl I}-band image for this
object reduces the constraints available for the fitting.  A19 is one
of two objects with evidence of a broad MgII emission line component
in its optical spectrum (Table~\ref{tab:ew}). The X-ray spectrum
is, however, fit with a power-law model and excess absorption 
(see section 2.5). Combined with its high $B$-band magnitude and
stellar spatial profile, we are most likely viewing the \lq naked'
quasar in this object. (The only other object showing a similar broad
component to the MgII emission line is A16, which also has little
X-ray absorption.)

The observed broadband spectrum shown in Figure~\ref{fig:a19sed} is
also consistent with a very young early type system before the Balmer break
sets in. In fact, the best photometric-redshift fit to the A19
magnitudes using galaxy templates, and restricting the fits to M$_{\rm
B}$ fainter than $-$25 gives such a young (0.26 Gyr) early-type
system, but at a redshift of z$_{\rm{phot}}=0.52$, different to z$_{\rm
spec}$.  A broken power-law continuum template (due to Granato, Danese
\& Franceschini 1997) gave a fit at z$_{\rm{phot}}=2.36$ with a
significantly worse $\chi^2_{\rm{reduced}}=9$, while a free fit with a
high signal-to-noise empirical QSO spectrum (from Francis et al. 1991)
gives z$_{\rm{phot}}=0.32$. All the fits are far from the inferred
z$_{\rm{spec}}$. This is primarily because the photometric-redshift
fitting technique relies on breaks in the broadband spectrum, which
this object does not possess. We do not think that the discrepancy in
redshifts could be due to contamination by strong line emission. To
demonstrate this, we fit the magnitudes by the empirical unobscured
QSO spectrum of Francis et al (1991), but now with its redshift fixed
at $z=z_{\rm{spec}}=1.675$. The fit is plausibly acceptable
(Fig~\ref{fig:a19sed}), except where it falls well below the observed
flux in the {\it R}-band. However, there are no obvious emission lines
between MgII and CIII] that could be boosting the {\it R} magnitude.


Finally, we also attempted to fit the colours and magnitudes of A19 by
combined quasar$+$host galaxy spectra, quantifying the amount of
dilution by comparing the equivalent width (EW) of the observed
emission lines (e.g. EW(MgII in A19)$=34\rm{\AA}$) to the EW found by
Francis et al. (1991) for an average unobscured quasar spectrum
(50$\rm{\AA}$ for MgII). This decrement can be interpreted as an
increment in the continuum due to the host galaxy contribution. An SED
was constructed with a typical AGN-power law contribution at 34/50
that of the total continuum at 2797$\rm{\AA}$, the rest-wavelength of
MgII. Four different SEDs were constructed with the four CWW host
galaxies models. However, the best-fit to A19 was a
powerlaw-plus-continuous-star-formation model at $z_{\rm{phot}}$=0.06
and $\chi^2_{\rm{reduced}}=10$, again far from $z_{\rm{spec}}$. We
also tried a reddened-AGN-spectrum-plus-galaxy spectrum, but again
obtained the wrong redshift. Thus we infer that if the photometry and
optical-line identifications are correct, the template of this galaxy
does not exist within our library, although the MgII
broad line most likely indicates a relatively-unobscured quasar. {\sl
I}-band photometry would be a first-step to constrain this object
more. 
We note that A19 appears to fit into the class of objects
identified by Maiolino et al (2001) which have blue optical colours,
and moderate X-ray absorption. The reduced absorption due to dust
compared to that expected from the gas column density suggests 
that such AGN have dust-to-gas ratios very different to Galactic
values. If this is the case for A19, then it is not surprising that
HYPERZ has difficulty in finding a solution. 

A16, the third object with a slight redshift mismatch, is similar to
A19. As already mentioned these are the two sources with good fits to
broad MgII$\lambda$2798 components. There is no break in the
broad-band optical/near-infrared SED of either, with
flux-per-unit-wavelength increasing monotonically towards the optical.
A16 is, in fact, best fit with a naked QSO spectrum at a redshift of
1.35, with the spectroscopic redshift of 1.6321 being encompassed
within the 90\% confidence interval of the HYPERZ fit. This is
consistent with the X-ray spectral analysis for this object, which
reveals little absorption (Table~\ref{tab:xsp}). Curiously, A16 and
A19 are also located close together on the sky, with a separation of
just over an arc-minute between them.

\subsection{X-ray spectra}

We attempted to extract and model the X-ray spectrum of those
serendipitous sources for which a redshift (either spectroscopic or a
photometric estimate) was available. The fitting was performed in
XSPEC (Arnaud 1996). In practice, we can extract a usable spectrum only
for the brighter sources (A6, A15, A16, A18, A19, A20, A24 and A26).
We were not able to fit A28, but note that most of the counts lie
between 0.7 and 0.8\keV. We fit the model to the data only between 0.3
-- 0.7\keV, and the spectra were grouped to have a minimum of 20
counts in each bin. In all cases we assumed the source to have a
power-law spectrum with possible intrinsic absorption, ie in excess of
the Galactic column density. In all cases we let both the power law
slope ($\Gamma$) and the excess $N_{\rm H}$ vary freely; if this fit
required a value of $\Gamma$ that was appreciably different from the
canonical value of $\Gamma=2$, we repeated the fit with $\Gamma$ fixed
at 2. Unless otherwise noted in Table~\ref{tab:xsp}, the data from
both observations were fit simultaneously. We present the results in
Table~\ref{tab:xsp}. Although A8 had too few counts to fit the spectra
properly, we can at least say that the spectra appear to be consistent
with no intrinsic absorption, for a power-law model with $\Gamma=2$
fixed.

\begin{figure}
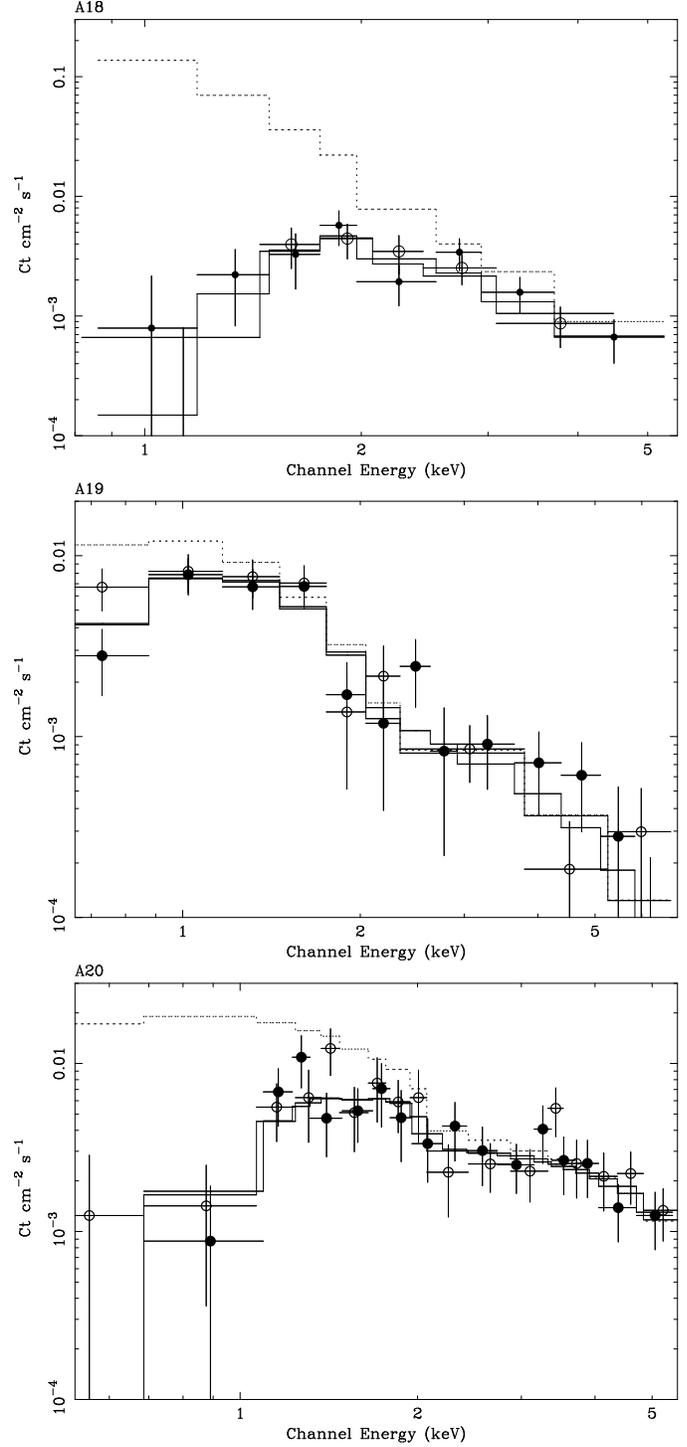

\vbox{
\psfig{figure=xspeca18.ps,width=0.5\textwidth,angle=270}
\psfig{figure=xspeca19.ps,width=0.5\textwidth,angle=270}
\psfig{figure=xspeca20.ps,width=0.5\textwidth,angle=270}}
\caption{\label{fig:a20xspec}
The X-ray spectra of A18, A19 and A20, showing the best-fit absorbed
power law models to the data (solid line), and the fit but now without the
intrinsic absorption (dotted line). The parameters of these best-fit
models are given in Table~\ref{tab:xsp}. The solid and open circle
markers indicate the spectrum extracted for the source from the two
different datasets. }
\end{figure}

\begin{figure}
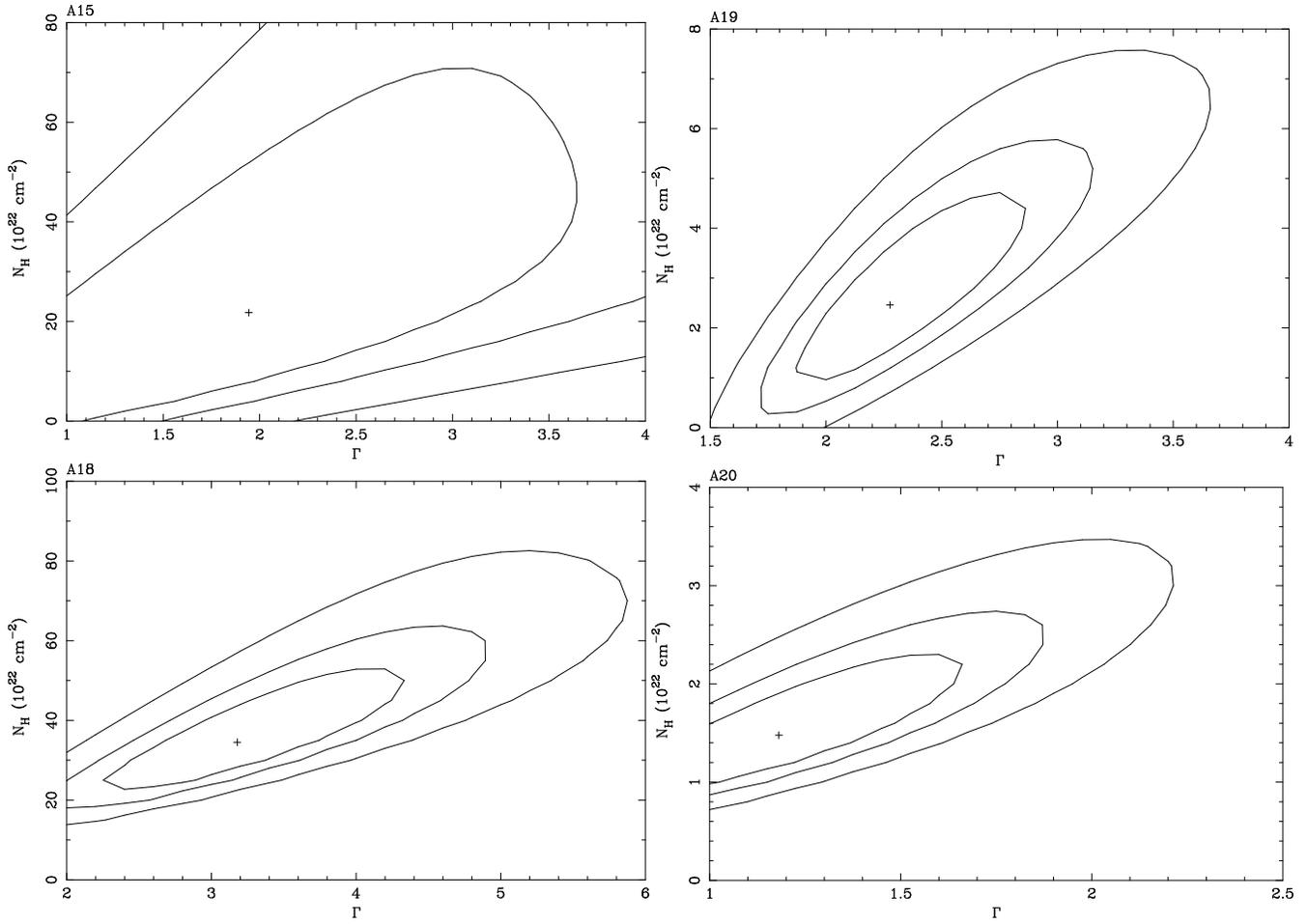

\hbox{
\vbox{
\psfig{figure=a15conf.ps,width=0.5\textwidth,angle=270}
\psfig{figure=a18conf.ps,width=0.5\textwidth,angle=270}
}\vbox{
\psfig{figure=a19conf.ps,width=0.5\textwidth,angle=270}
\psfig{figure=a20conf.ps,width=0.5\textwidth,angle=270}}}
\caption{\label{fig:confcont} 
The confidence contours on the parameters $\Gamma$ and intrinsic absorption N$_{\rm H}$
for the X-ray spectral fits to A15, A18, A19 and A20. 
}
\end{figure}

The best fits to the spectra of A18, A19 and A20 are shown in
Fig~\ref{fig:a20xspec}, demonstrating that A18 and A20 require
substantial amounts of excess absorption, whereas A19 does not. The
contours of confidence for the values of freely fitting parameters
N$_{\rm H}$ and $\Gamma$ in the fits to A15, A18, A19 and A20 spectra
are shown in Fig~\ref{fig:confcont}. These show that the conclusion of
excess absorption is robust for A18, A19 and A20 (although less so for
A15) for all reasonable values of $\Gamma$.
Note that the intrinsic absorption $N_{\rm H}$ inferred from the X-ray spectra 
is in all cases more than that implied by the moderate amounts of
reddening A$_{\rm V}$ (Table~\ref{tab:zphot}) found from the photometric
redshift fitting in the previous section. However, 
 any reddening of the host galaxy is due to dust within the galaxy, whereas
the line of sight to an active nucleus is expected to pass through
additional absorption much closer in to the central engine. As discussed in the previous section, one source, A19, shows evidence for a dust-to-gas ratio other than Galactic.

It is also possible to examine the faint sources which do not have
enough counts to justify spectral fitting. We used XSPEC to predict
the soft-to-hard (S/H) ratio that would be observed (on the I3, S2, S3
and S4 chips) for an absorbed (intrinsic + Galactic) power-law source
placed at a range of redshifts with a variety of intrinsic column
densities (cf. Table~4 in Crawford et al. 2001. Note that this table
assumes zero Galactic absorption). Next, the observed S/H ratios
(Table~1) for sources with a redshift determination were compared with
the predicted values, from the same chip, in order to constrain the
intrinsic column densities. These are listed in
Table~\ref{tab:shnh}. Predictions were made for powerlaw-indices
$\Gamma$ of 1.4 and 2. Note that the maximum observable S/H ratio for
a fixed photon-index will be from an intrinsically-unabsorbed
source. Any non-zero intrinsic absorption could only deplete the soft
counts and decrease the S/H ratio. This is basis of the predictions of
larger $\Gamma$ in the last column of Table~\ref{tab:shnh} for the
very soft sources (with observed S/H greater than the prediced maximum
for $\Gamma=2$ and for $\Gamma=1.4$). It is not possible to constrain
the intrinsic absorption in these cases, since larger photon-indices
could also imply larger column densities.

\onecolumn
\twocolumn

\section{Summary of results and Discussion}


We have detected 31 serendipitous sources in the field of A\,2390. The
majority of these (18 in total) were detected on the ACIS-S3 chip,
which contains the pointing target of the observations, A\,2390. 
We do not consider that our results are enhanced by weak gravitational
lensing by the cluster. The degradation of the PSF and vignetting make
the other chips less sensitive to faint sources. Only
three of the sources are sufficiently close to the line of sight
through the centre of the cluster potential to be
significantly magnified by gravitational lensing. These are A15, A17
and A18, which are magnified by 7.8, 2.8 and 2.1 respectively (Cowie
et al 2001). None of the sources appears significantly more extended
than the PSF expected at that position on the detectors, and are hence
all consistent with being point sources. About 50 per cent of the
sample have soft X-ray spectra (S/H$>3$), and about one-third of the
sources show hard spectra (ie S/H$<2$). The six sources with
the hardest colours (S/H$<1.5$) are A2, A15, A18, A20, A24 and A30.
Eight of the sources showed evidence for variability on an
eleven-month timescale. 

We have fitted an absorbed power-law model to simple X-ray spectra of
the eight sources with sufficient counts and a known redshift. The
unabsorbed (de-magnified) 2-10\keV\ luminosity ranges from 0.1 to
30$\times10^{44}$\ergps. The intrinsic absorption required varies from
none (in two cases) to 35$\times10^{22}$\pcmsq. Most of these sources
are thus in the Compton-thin regime. We note that the two sources for
which the X-ray spectrum indicates the most intrinsic absorption
(N$_{\rm H}>10^{23}$\pcmsq) are those -- A15 and A18 -- that are
magnified from gravitational lensing by the foreground cluster A\,2390.
Without the presence of the foreground cluster we would not have been
able to observe these sources so clearly.  We have found no sources in
this field which are clearly Compton-thick (i.e. have an absorbing
column $>10^{24}\pcmsq$). The lack of such sources is, however,
consistent with predictions of the observable density of Compton-thick
objects at this flux level (Wilman \& Fabian 1999). As well as its
high intrinsic column, A18 also has an intrinsic luminosity of
$>10^{45}$\ergps (correcting for the gravitational lensing magnification);
thus it is a good candidate for an X-ray Type~II quasar.

\begin{figure}
\psfig{figure=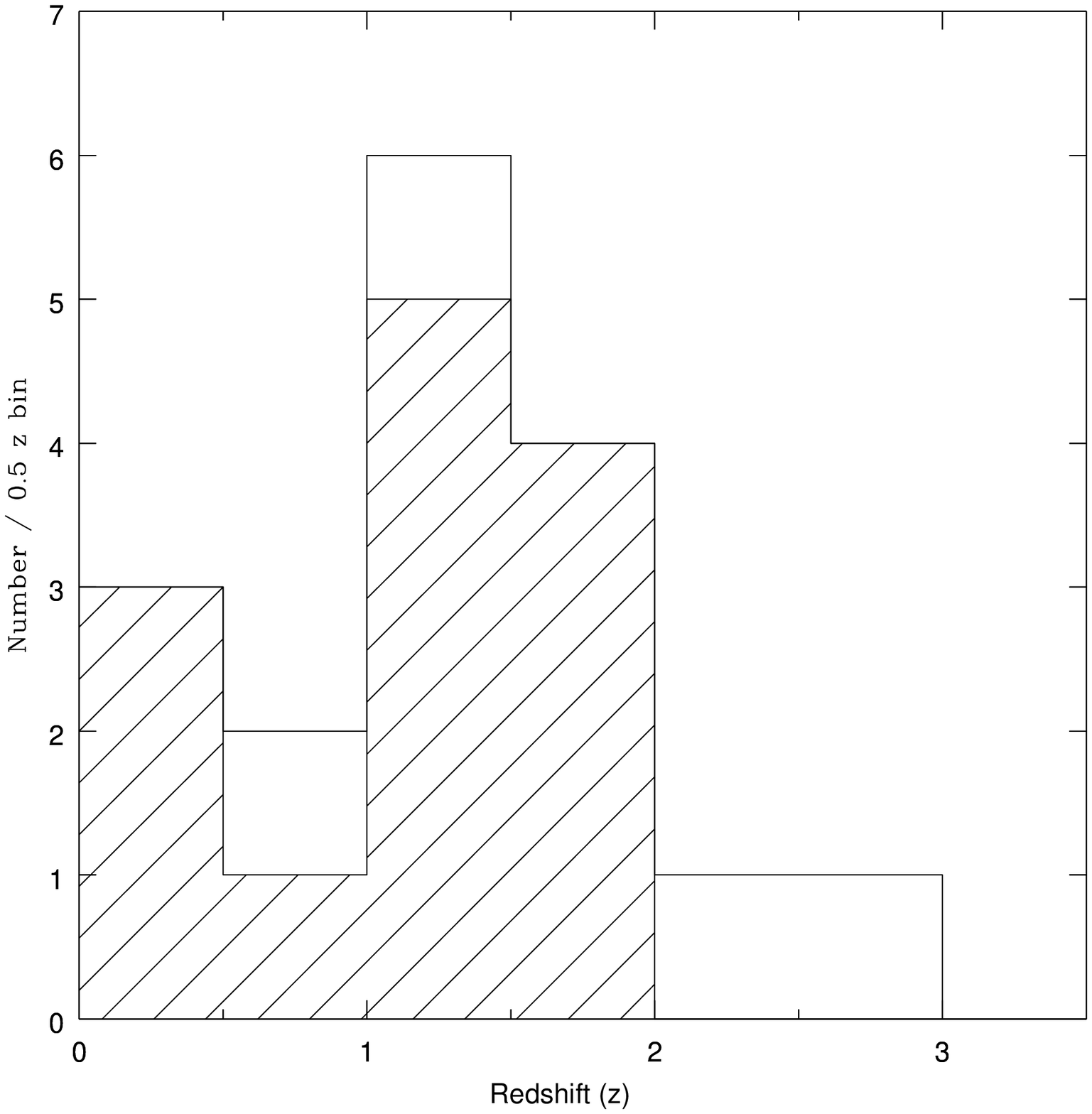,width=0.5\textwidth,angle=0}
\caption{\label{fig:zhist} The redshift distribution, where known, of
the serendipitous sources in the field of A\,2390. The shaded boxes
indicate those sources that have a spectroscopic redshift
rather than a photometric redshift. }
\end{figure}

We have clear optical identifications for 23 of the sources, and
 imaged twelve in the near-IR, choosing preferentially the
 optically-faint harder ones. All the sources were detected in $J$,
 $H$ and $K$, including one source (A15) that was not detected down to
 faint optical levels. Optical spectra were taken for fifteen of the sources
 (including two objects which had stellar spectra; note also that
 the magnitudes of A9 and its stellar profile in the DSS strongly
 suggest that this source is also a star). We have obtained a new
 spectroscopic redshift for 10 sources (3 others were already in the
 literature), and estimated a photometric redshift for a further four
 sources (one of which is in agreement with a previously published
 photometric estimate). We tested the results of photometric redshift
 fitting for those sources for which we also had optical spectra, and
 found reasonably good agreement. We show the redshift distribution
 for all the sources in Fig~\ref{fig:zhist}. The four very hard X-ray
 sources for which we have redshifts span the whole redshift range
 observed, at 0.214, 0.305, 1.467 and 2.78. The remainder of the
 optical spectra showing emission lines are of softer sources. We have
 photometric redshifts for a further 3 hard sources.

\begin{figure}
\psfig{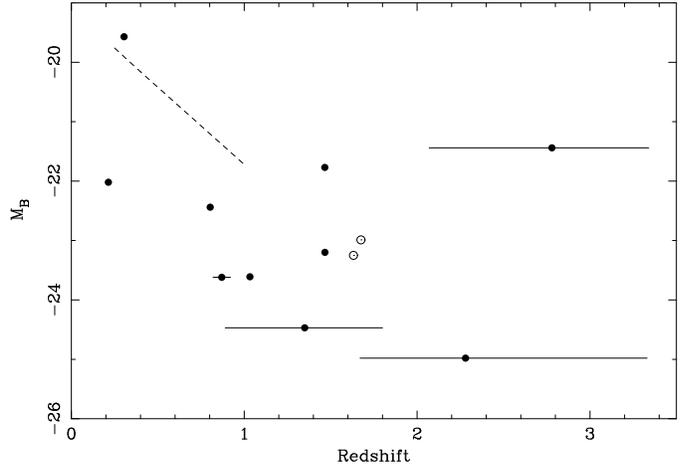}
\caption{\label{fig:magz} The absolute M$_{\rm B}$ of the host
galaxies from the HYPERZ fitting plotted against redshift. The dashed
line shows the magnitude-redshift relation of blue field galaxies out
to $z\sim1$ from Lilly et al (1995). A16 and A19 are marked by open
circle symbols, as their light is thought to be dominated by a quasar
component. }
\end{figure}

The $K$ magnitude and the redshift $z$ for all of our sources follow
the relationship known for radio galaxies (Fig~\ref{fig:kz}; Eales et
al 1993), even though none of our serendipitous sources are coincident
with radio detections listed in the NASA/IPAC Extragalactic Database
(NED; this area of sky is not covered by the FIRST catalogue).
This results contrasts with the tendency for the {\sl
Chandra} sources found in the ELIAS survey area to lie well above the $K-z$
relation  (Willott et al 2001). 
The
{\sl B} magnitude of the host galaxies inferred from the photometric
redshift fitting increases with redshift (Fig~\ref{fig:magz}).
Comparison of our magnitudes to the the magnitude-redshift
distribution of the blue field galaxy out to $z\sim1$ (from Lilly et
al 1995), indicates that the majority of our sources lie in brighter,
and presumably more massive galaxies than the general field
population.


All the sources (except for A15) with spectroscopic redshifts show
optical emission-line spectra, indicative (in combination with their
point-source appearance in the optical/near-IR bands and the large
fraction showing X-ray variability) of AGN-type
behaviour. Seven of the nine sources at high enough redshift (and with
optical spectra) show the MgII emission line. We note that A18 -- one
of the two sources with large amounts of intrinsic absorption -- shows
no sign of MgII emission (although its expected position is right on
the edge of the observed spectrum). The other heavily absorbed source,
A15, shows no sign of CIII]$\lambda$1909, which should be observed at
$\sim7180$\AA\ if the source is at the photometric estimated redshift
of 2.78 (Fig~\ref{fig:nozsp}).  We do not have an X-ray spectrum for
the other source that does not show MgII emission (A8), although we
note that it is a soft X-ray source (Table~1). There does not seem to
be a broad component to the Balmer emission lines in A20 or in A24,
which are two X-ray-hard sources, at least one of these being a narrow-line Seyfert.

Though we do not enough sources for clustering statistics, we note
that at least two sources (A17 and A18) lie at the same redshift, with
a separation of about 40 arcsec. 
Two other sources (A16 and A19) -- both showing evidence of harbouring
relatively-unobscured AGN -- are also separated by about an arcmin, at
redshifts of 1.632 and 1.675 respectively. 

\begin{figure}
\psfig{figure=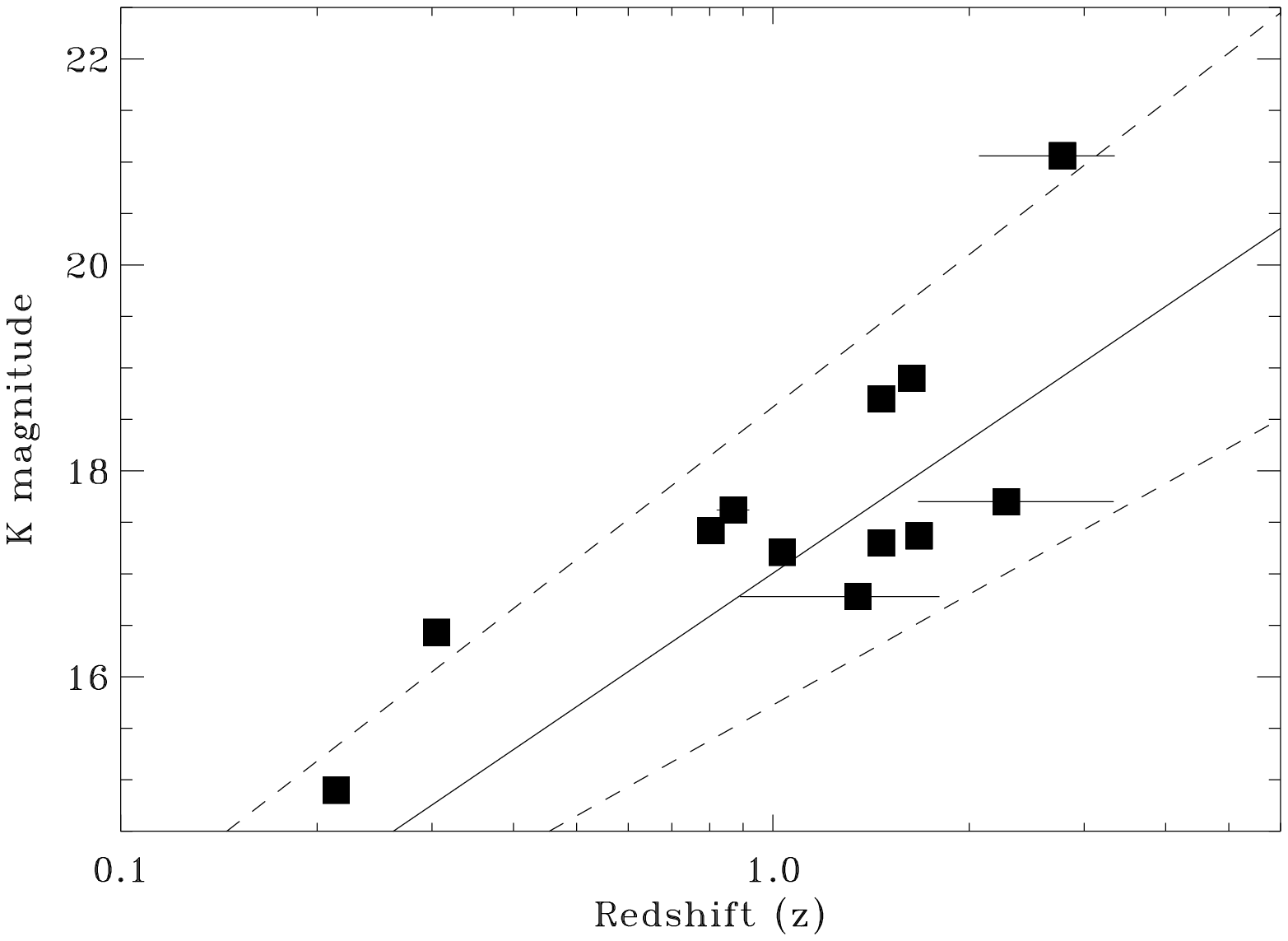,width=0.5\textwidth,angle=0}
\caption{\label{fig:kz} The $K$-$z$ relation for our sources. The
solid line shows the $K$-$z$ relation for radio galaxies, and the
dotted lines indicate the scatter of the radio galaxies about this
line (taken from Eales et al 1993). The K-magnitudes of A15, A17 and
A18 have been corrected for the magnification due to gravitational
lensing. }
\end{figure}

\section{Conclusions}
Optical and near-infrared follow-up work on two $\sim$10~ks {\sl
Chandra} exposures of the A\,2390 field resulted in 23 out of 31
detected sources being identified. The 0.5--7\keV\ fluxes of these
sources range from just below, to a few times, $10^{-14}\ergpcmsqps$.
Two of the sources are stars and 13 have optical emission-line
spectroscopic redshifts ranging from 0.2--1.7. One source, A10 at
$z=0.2242$, probably lies within the cluster A\,2390. Apart from the
two lowest redshift non-cluster objects, the rest with optical
emission lines have soft X-ray spectra. We have JHK photometry of 12
sources which enabled us to obtain photometric redshifts for a further
4 sources. We find good agreement between photometric and
spectroscopic redshifts where both are available. Of the 12 sources
without either optical spectroscopy or near-infrared imaging, about
half are probably hard (S/H$<2$). 4 of this hard group are
undetectable on the DSS and 4 have no identification.

We have obtained an insight into the properties of the harder, fainter
population from the 3 gravitationally-lensed sources (A15, A17 and
A18). All would have been detected in X-rays without the lensing, A15
and A17 would have been just detected and A18 well detected. A15 and
A18 have hard X-ray spectra. The magnification has made the
near-infrared work easier and Cowie et al (2001) have obtained
(similar) infrared spectroscopic redshifts ($z=1.47$) for both A17 and
A18 and a photometric redshift ($z\sim2.6$ confirmed here) for A15. A15
and A18 appear to be genuine obscured quasars with intrinsic 2--10~keV
luminosities of $2\times 10^{44}\ergps$ and $2\times 10^{45}\ergps$
respectively. They were both also detected in the mid-infrared by ISO.
Model spectral energy distributions for the $z\sim2.6$ source indicate
that the reprocessed radiation is emitted by hot (1500\K) or warm
dust, peaking at 100-200$\mu$m.

The results of our modest sample are consistent with, and extend,
previous work on deeper fields (e.g. Barger et al 2001). We find that
i) most of the sources at 0.5--7\keV\ fluxes around
$10^{-14}\ergpcmsqps$ can be optically identified, although deep
optical and infrared imaging is required; ii) about one half of the
sources, mostly the X-ray softer ones, have optical emission lines
with which spectroscopic redshifts can be obtained; iii) photometric
redshifts work well and enable some more redshifts to be obtained,
particularly of harder sources; iv) about one third of the sources are
both optically faint and hard. These last sources are likely to be
important for producing the harder parts of the XRB, which peaks in
$\nu I_{\nu}$ at $\sim 30\keV$, although the class of faint
Compton-thick sources which should dominate at the peak may not yet
have been detected. If A15 is typical of the optically-faint hard
sources, then they are obscured quasars in early-type bulges at
redshifts of 2--3, or possibly higher. The absorbed X-ray and UV
emission emerges in the mid-infrared.

\section{Acknowledgements}

We are grateful to the {\sl Chandra} project for the X-ray data, and
to M. Bolzonella J.-M. Miralles and R. Pello for making HYPERZ
available. We thank Stefano Ettori for the use of his software in
reducing the X-ray data, and Andrew Firth for his help with HYPERZ.
CSC and ACF thank the Royal Society, PG thanks the Isaac Newton Trust,
the Overseas Research Trust and Fitzwilliam College Trust Fund, and
RJW the PPARC for financial support. Support for AJB was provided by
NASA through the Hubble Fellowship grant HF-01117.01-A awarded by the
Space Telescope Science Institute, which is operated by the
Association of Universities for Research in Astronomy, Inc., for NASA
under contract NAS 5-26555.  AJB and LLC acknowledge support from NSF
through grants AST-0084847 and AST-0084816, respectively.

The United Kingdom Infrared Telescope is operated by the Joint
Astronomy Centre on behalf of the U.K. Particle Physics and Astronomy
Research Council. This research made use of data obtained from the
Isaac Newton Group Archive at the UK Astronomy Data Centre, Cambridge,
and the Canadian Astronomy Data Center, which is operated by the
Dominion Astrophysical Observatory for the National Research Council
of Canada's Herzberg Institute of Astrophysics. This research has also
made use of the NASA/IPAC Extragalactic Database (NED), and the
Digitized Sky Surveys which were produced at the Space Telescope
Science Institute under U.S. Government grant NAG W-2166.

{}

\onecolumn
\begin{figure*}
\addtocounter{figure}{-1}
\psfig{figure=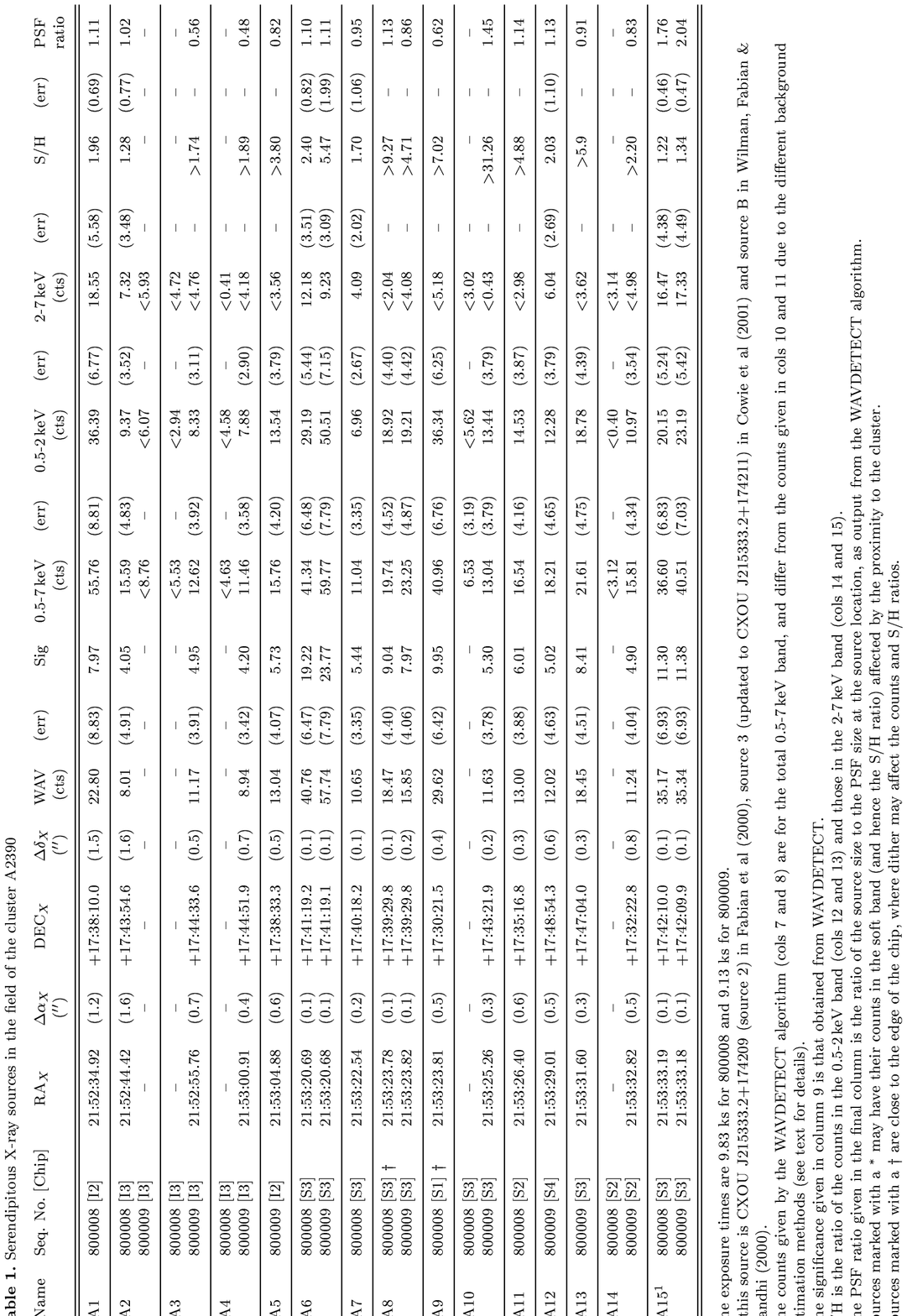,width=1.0\textwidth,angle=180}
\end{figure*}
\begin{figure*}
\addtocounter{figure}{-1}
\psfig{figure=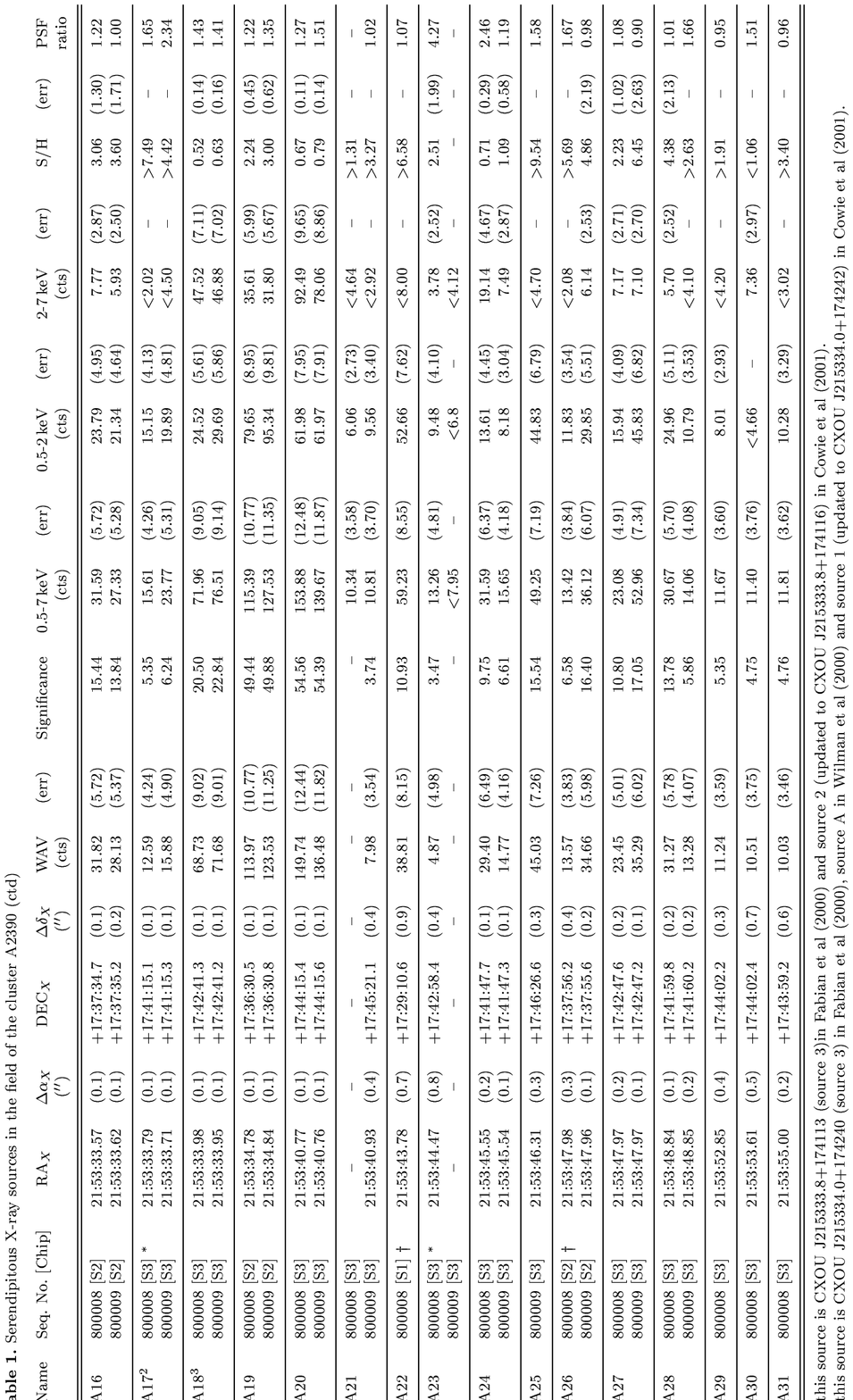,width=1.0\textwidth,angle=180}
\end{figure*}

\addtocounter{table}{+1}
\onecolumn
\begin{table}
\caption{A\,2390 Observations\label{tab:obs}}
\begin{tabular}{lclclcc}
              && & & &  \\
Band & Date & Telescope \& Instrument & Plate Scale & Filters & Seeing & Typical Exposure\\
     &      &                         & (arcsec/pixel) &      & (arcsec) & (seconds)\\
\hline 
B & 1994 Dec 08 & WHT Prime Focus &  0.421 & Kitt Peak 1 & 3.8 & 500\\
R & 1994 Jun 06 & INT Prime Focus &  0.590 & Kitt Peak 3 & 1.6 & 600\\
I & 1994 Dec 08 & WHT Prime Focus &  0.421 & Kitt Peak 1 & 1.4 & 300\\
I & 1990 Oct 16 & CFHT focam      &  0.205 & CFHT \#1809 & 0.7 & 700\\
J, H, K & 2000 Aug 10/11 &  UKIRT UFTI & 0.091 & J98, H98, K98 & 0.6 & 540\\
\hline 
\end{tabular}
~~~\\
\end{table}

\begin{table}
\caption{A\,2390 Photometry \label{tab:mags}}
\begin{tabular}{lcccccccccc}
 Object  & RA$_{\rm opt}$ & Dec$_{\rm opt}$ &B & R & I & J & H & K & Aperture diameter\\
  & & & & & & & & & (arcsec)\\
\hline
A3 & 21:52:55.80    & 17:44:34.1 & $>$22.5$^D$ & $>$21.0$^D$  & -- & -- & -- & -- & \\
A4 & 21:53:00.95    & 17:44:52.4 & 22.0$\pm$0.5$^D$ & $>$21.0$^D$  & --  & -- & -- & -- & \\
A5 & 21:53:04.92    & 17:38:33.8 & 16.89$^A$& 14.80$^A$& --  & -- & -- & -- & \\
A6 & 21:53:20.72    & 17:41:19.6 & $>$22.5$^D$ &21.96$\pm$0.10 & --  & -- & -- & -- & 4.5\\
A7 & -- & -- & $>$22.5$^D$ & $>$23.40 & $>$22.25 & -- & -- & -- & 4.5\\
A8 & 21:53:23.86    & 17:39:30.3 & 23.69$\pm$0.37 & 21.84$\pm$0.08 & 20.24$\pm$0.06 & 19.50$\pm$0.06 & 18.38$\pm$0.04 & 17.42$\pm$0.03 &  4.5\\
A9$^A$ & 21:53:23.68   & 17:30:19.9  &13.31$^A$ &12.05$^A$ & --  & -- & -- & -- & --\\
A10 & 21:53:25.30    & 17:43:22.3 & 22.39$\pm$0.09 & 19.91$\pm$0.03 & 19.01$\pm$0.03 & --& -- & -- & 4.5 \\
A11$^D$ & 21:53:26.47 & 17:35:17.4  & 22.0$\pm$0.5$^D$ & $>$21.0$^D$  & -- & -- & -- & -- & \\
A12$^P$ & 21:53:29.05 & 17:48:54.7  &   $>$22.5$^D$ & $>$21.0$^D$  &  --  & 19.48$\pm$0.05 & 18.35$\pm$0.03 & 17.70$\pm$0.03 & 4.5\\
A14 & 21:53:32.86    & 17:32:23.8 & $>$22.5$^D$ &  $>$21.0$^D$ &  --  & 18.32$\pm$0.03 & 17.59$\pm$0.03 & 16.78$\pm$0.02 & 4.5\\
A15$^S$ & 21:53:33.22    & 17:42:10.4 & $>$23.72 & $>$23.25 & $>$22.88$\ddag$ & $>$21.71 & 20.07$\pm$0.06 & 18.83$\pm$0.04 &  4.5\\
A16 & 21:53:33.65    & 17:37:35.6 & 22.0$\pm$0.5$^D$ & 20.98$\pm$0.06 &  --  & 19.82$\pm$0.07 & 19.55$\pm$0.10 & 18.90$\pm$0.10 &  2.7\\
A17 & 21:53:33.75    & 17:41:15.8 & $>$23.66 & 22.99$\pm$0.24 & 21.11$\pm$0.10$\ddag$ & 19.38$\pm$0.03 & 18.59$\pm$0.03 & 17.58$\pm$0.02 &  2.7\\
A18 & 21:53:33.99    & 17:42:41.6 & $>$23.74 & 23.86$\pm$0.37 & 21.13$\pm$0.11$\ddag$ & 18.65$\pm$0.03 & 17.57$\pm$0.02 & 16.49$\pm$0.02 &  4.5\\
A19 & 21:53:34.88    & 17:36:31.7 & 20.50$\pm$0.30$^A$ & 19.35$\pm$0.03 &  --  & 18.74$\pm$0.03 & 17.80$\pm$0.03 & 17.37$\pm$0.03 &  4.5\\
A20 & 21:53:40.80    & 17:44:16.0 & 21.90$\pm$0.07 & 20.18$\pm$0.03 & 19.38$\pm$0.02 & 18.22$\pm$0.02 & 17.64$\pm$0.02 & 16.43$\pm$0.02 &  4.5\\
A21 & 21:53:40.97    & 17:45:21.6 & $>$22.5$^D$ & $>$23.34 & --  & -- & -- & -- & 4.5\\
A22$^D$ & 21:53:43.87 & 17:29:12.3  & 20.48$^A$& 20.5$\pm$0.5$^D$  & -- & -- & -- & -- & \\
A23 & -- & -- & $>$23.67 & $>$23.89 & $>$22.49 & -- & -- & -- & 2.7\\
A24 & 21:53:45.58    & 17:41:47.8 & 20.26$\pm$0.04 & 18.24$\pm$0.02 &
17.41$\pm$0.03 & 16.31$\pm$0.02 & 15.63$\pm$0.02 & 14.90$\pm$0.02 &
4.5\\
A25 & 21:53:46.35    & 17:46:27.0 & 20.20$^A$& 17.76$^A$&-- & -- & -- & -- & \\
A26 & 21:53:48.00    & 17:37:56.1 & 20.10$^A$& 19.85$\pm$0.03 & -- & -- & -- & -- & 4.5\\
A27 & 21:53:47.98    & 17:42:48.0 & 21.97$\pm$0.09 & 20.91$\pm$0.05 & 19.22$\pm$0.04 & 18.97$\pm$0.04 & 18.42$\pm$0.04 & 17.62$\pm$0.03 & 2.7\\
A28 & 21:53:48.87    & 17:42:00.0 & 22.20$\pm$0.10 & 21.30$\pm$0.08 & 20.58$\pm$0.05 & 18.98$\pm$0.03 & 18.19$\pm$0.03 & 17.21$\pm$0.02  &  4.5\\
\hline 
\end{tabular}
~~~\par

The following sources are not included in the table as they are not in
the field of view of the archival optical datasets, and there is no
obvious id in either the DSS B or R images: A1 A2, A29, A30 and A31. \\
The RA$_{\rm opt}$ and Dec$_{\rm opt}$ listed in columns 2 and 3 are
from Barger (private communication); except for those sources whose
name is marked by $^A$ or $^D$ where the position is from the
APM or DSS data respectively. \\
In columns 4 and 5, a magnitude (or upper limit) marked by a $^D$ or $^A$ 
was estimated from the DSS (second-generation) image, or from the Palomar Observatory Sky Survey as
scanned from the APM at Cambridge, 
as the source was not in the field of view
of the archival optical data. The remaining magnitudes are either from
the INT ($R$-band), or the WHT ($I$- and $B$-band) image, except for
the 
three $I$-band magnitudes marked $\ddag$, which were measured from the
deeper CFHT image.\\
The source marked $^P$ has a systematic photometric uncertainty of 14\% in J, 12\% in H and 8\% in K. The value stated has been calibrated against all standards observed through the night - see text.\\
The aperture size stated in the last column is for the infra-red data. For the optical data, the seeing-corrected apertures were typically larger.\\
The source marked $^S$ also has {\sl ISO} 6.7- and 15- ${\rm{\mu}}$m detections of 110 and 350 
${\rm{\mu}}$Jy respectively (L\'emonon et al 1998). These translate to magnitudes of 14.8 and 12.7.\\

\end{table}

\begin{table}
\caption{Equivalent widths of emission lines, in \AA. \label{tab:ew}}
\begin{tabular}{llllllllll}
Source & CIII]$\lambda1909$ &MgII$\lambda$2798 & FWHM of Mg II & [OII]$\lambda$3727& H$\beta$& [OIII]$\lambda$5007& H$\alpha$& [NII]$\lambda$6584& [SII]$\lambda$6731 \\
& & & (\kmps) & & & & & & \\
\hline 
A4 & --  &  21.2$\pm{1.0}$ & 4202$\pm$226 & 1.7$_{-1.1}^{+1.2}$&  -- & --& --& -- &--\\
A6 & --   &  35.6$_{-2.2}^{+2.3}$ & 6995$\pm$465 & 9.1$_{-1.4}^{+1.5}$ & -- & --& --& -- &--\\
A8 & --   &  $<$2.5 & -- & 24.8$\pm{0.7}$& 15.2$_{-1.2}^{+1.1}$&27.9$\pm{2.0}$&  --& -- & --\\
A10 &--   &  --   & -- &  --   &  --    & --   &  0.9$\pm{0.5}$& 5.4$_{-0.5}^{+0.6}$ & -- \\
A13 &33.6 $_{-1.7}^{+1.8}$& 49.6$_{-2.3}^{+2.4}$ &8455$\pm$139 & -- & --& --& -- &--&
--\\
A16 &25.2$_{-2.0}^{+2.2}$ & 33.9$_{-2.4}^{+2.5}$& 8203$\pm$453$^\dag$ & -- & --& --& -- &
--& --\\ 
A18 &--   &  --   & -- &  13.5$_{-2.7}^{+3.1}$ &  --& --& -- &--& --\\
A19 &30.7$\pm{1.4}$ & 45.6$\pm{1.9}$ & 8271$\pm$403$^\dag$ & -- & --& --& -- &
-- & --\\
A20 &--   &  --   & -- &  163.6$\pm{6.1}$ & 71.1$_{-7.3}^{+7.2}$  & 711.8$_{-25.7}^{+26.1}$& 249.3$_{-11.3}^{+11.7}$& 53.5$_{-6.6}^{+6.5}$& 10.7$\pm{6.2}$ \\
A24 &-- & -- & -- & --   &  --   & 35.1$\pm{1.6}$& 18.7$_{-0.9}^{+1.0}$&32.2$_{-1.0}^{+1.5}$& 2.7$\pm{0.8}$ \\
A26& 24.2$\pm{1.2}$& 23.0$\pm{0.7}$ & 3825$\pm$126 & -- & --& --& -- & --& --\\
A28 &-- & 79.2$\pm{3.7}$ & 6089$\pm$277 &  -- & --& --& -- & --& -- \\
\hline
\end{tabular}
~~~\\
$^\dag$ For A16 and A19, the MgII line is fit better by a double
Gaussian, whereas the equivalent width and velocity width listed in
the table are from a single
Gaussian fit. In A16 these two components have velocity widths of 
2020$\pm$239 and 10593$\pm$516\kmps, with the broader component having 
$\sim6$ times the intensity of the narrow line. 
In A19 the  two 
components have velocity width 3548$\pm$503 and 12933$\pm$679\kmps, and the 
broader component is $\sim$7 times the intensity of the narrow line. 
\end{table}

\begin{table}
\caption{A\,2390 Redshifts \label{tab:zphot}}
\begin{tabular}{llcclcccc}
Source & $z_{\rm spectroscopic}$& $z_{\rm photometric}$ & Reduced $\chi^2$ (d.o.f.)  & Galaxy Type & Age
& $A_{V}$ & M$_B$ & $z_{\rm secondary}$  \\
             & &(90\% confidence)& & &(Gyr)& & & \\
\hline
A4 & 1.2172 & -- & -- & -- & -- & -- & -- \\
A6 & 1.0690  & -- & -- & -- & -- & -- & -- \\
A8  & 0.8030 & 0.79 (0.75,0.82) & 1.98 (5)  & Burst & 0.36 & 0.60 & --22.44 & 4.33 \\
A10 & 0.2242  & -- & -- & -- & -- & -- & -- \\
A12$^P$  & - & 2.28 (1.67,3.33) & 0.04 (4) & Burst  & 0.26 & 0.30 & --24.98 & 0.63  \\
A13 & 1.5933  & -- & -- & -- & -- & -- & -- \\
A14  & - & 1.35 (0.89,1.80) & 0.04 (4) & Burst & 0.72 & 0.00 & --24.47 & 5.29 \\
A15  & 2.6 $^{C3}$  &2.78 (2.07,3.34) & 0.19 (6) & E & 1.43 &
1.80 & --21.44$^\ddag$ & 3.16 \\
A16  & 1.6321 & 1.35 (1.17,2.06) & 1.08 (4) & QSO & 0.04$^\dag$ & 0.30
& --23.25 & 0.36  \\
A17  & 1.466 $^{C2}$ &1.11 (0.99,1.15) & 0.39 (5) & Burst & 0.51 &
0.60 & --21.77$^\ddag$ & 4.63 \\
A18  & 1.467 $^{C1}$  & 1.45 (1.31,1.53) & 2.02 (5) & Burst & 2.60
& 0.00 & --23.20$^\ddag$ & 5.72 \\
A19  & 1.6750 & 0.52 (0.42,0.60) & 2.02 (4) & E  & 0.26 & 0.00
& --22.99 & 3.09  \\
A20  & 0.305 & 0.22 (0.18,0.38) & 1.40 (5) & CWW Scd with AGN &
2.60$^{\dag}$ & 0.60 & --19.57 & 0.33  \\
A24$^*$  & 0.214 & 0.26 (0.15,0.52) & 0.42 (5) & E & 4.5  & 0.00 & --22.02 & 0.47    \\
A26 & 1.5187  & -- & -- & -- & -- & -- & -- \\
A27  &-- & 0.87 (0.82,0.92) & 5.15 (5) & Burst & 0.18 & 0.00 & --23.62 & 0.54  \\
A28  & 1.0330 & 1.21 (1.15,1.33) & 0.19 (5) & S0 & 0.72 & 0.90 & --23.61 & 1.63 \\
\hline 
\end{tabular}
~~~\par 
M$_B$ in column 8 is the absolute Vega Magnitude in the {\sl
B} Bessell filter. Those sources with a magnitude marked by a $\ddag$
have been corrected for the magnification due to gravitational
lensing. \\ 
$^{C1}$ This is a spectroscopic redshift from Cowie et al
(2001). \\ 
$^{C2}$ This is a spectroscopic redshift from Cowie et al
(2001) based on the single identification of a line, but also
consistent with their photometric redshift estimate of 1.5$\pm0.2$. A
different identification would imply a redshift of 2.317 instead. \\
$^{C3}$ This is a photometric redshift of 2.6$_{-0.2}^{+0.1}$ from
Cowie et al (2001). \\ 
$^P$ There is a high degree of degeneracy
between this redshift estimate and the secondary solution, with small
changes in the photometry. See text for details.\\ 
$^*$ This source is
identified as an Sbc galaxy with a redshift of $z=0.21501$ by Yee et
al (1991). A fit to an Sbc template gives $z_{\rm phot}=0.3$.\\ 
$\dag$
This is a Coleman et al (1980; CWW), observed, fixed-age SED. The age
has been determined by associating with a Bruzual-Charlot synthetic
spectrum with redshift and reddening fixed to the values stated.
\end{table}

\begin{table}
\caption{ Fits to the X-ray spectra of the brighter sources \label{tab:xsp}}
\begin{tabular}{lcrrrl}
Source & $\Gamma$ & $\Delta N_{\rm H}$ &  Reduced $\chi^2$ (d.o.f.) &
L(2-10\keV) & notes\\
       &          & ($10^{22}$\pcmsq) & & ($10^{44}$\ergps) &   \\
\hline 
A6 $\dag$ & 2.1$^{+1.0}_{-0.6}$ & $<$0.61 & 0.77 (13) & 9.2 \\ 
A15    & 1.7$^{+2.0}_{-1.4}$ & 17.5$^{+60}_{-17.5}$ & 0.65 (9) & 13.0$^*$\\  
       & [2]                 & 21.9$^{+36}_{-14.6}$ &  0.60 (10) & 17.0$^*$ \\
A16    & 1.3$^{+2.6}_{-1.3}$ & 0.2$^{+12.8}_{-0.16}$ & 0.96 (6) & 2.9\\
       & [2]   & 3.8$^{+8.0}_{-3.8}$ & 0.92 (7) & 3.9 \\
A18    & 3.2$^{+1.3}_{-0.5}$ & $34.7^{+20}_{-14}$ & 0.57 (11) & 60.0$^*$ \\ 
       & [2]                 & 20.4$^{+7.1}_{-5.2}$ & 0.81 (12) & 20.0$^*$ \\
A19    & 2.31$^{+0.5}_{-0.4}$ & 2.6$^{+2.1}_{-1.6}$ & 0.91 (19) & 13 \\
A20    & 1.2$^{+0.2}_{-0.3}$ & 1.5$^{+0.9}_{-0.6}$ & 0.53 (30) & 1.0 \\
       & [2] & $2.7^{+0.7}_{-0.6}$ & 0.71 (31) & 1.0 \\
A24    & 1.3$^{+3.8}_{-1.7}$ & 1.8$^{+4.9}_{-1.8}$ &  0.60 (8) &0.1 &
fit to 800008 data only \\
       & [2] & 2.9$^{+2.6}_{-1.4}$ & 0.57 (9) & 0.1 & fit to 800008 data only \\
A26    & 4.2$^{+5.8}_{-2.3}$ & 3.9$^{+12}_{-3.9}$ & 0.92 (2) & 4.8 & fit to 800009 data only \\
       & [2] & $<1.9$ & 1.42 (3) & 3.0 & fit to 800009 data only \\
\hline
\end{tabular}
~~~~~~\\
L(2-10\keV) is the unabsorbed 2-10\keV\ luminosity. \\
A square bracket around the value of $\Gamma$ in column 2 denotes that
the parameter was fixed at this value. \\
$\dag$ The count rates for A6 (see Table~1) suggest the
source  is softer in the 800009 observation. Separate spectral fits to
each spectrum suggest that this may be due to a change in N$_{\rm H}$,
but since the errors on the separate fits overlap we present the
results here only for joint fitting. \\
$^*$ Sources A15 and A18 lie behind the cluster, and the 
luminosities listed are not corrected for the magnification factors of 7.8 and 2.1
respectively (Cowie et al 2001). \\
\end{table}

\begin{table}
\caption{ Absorbed-powerlaw-model-predicted intrinsic columns or photon indices from the observed S/H ratio for sources with insufficient counts for spectral fitting \label{tab:shnh}}
\begin{tabular}{lllr}
Source & Redshift & S/H ratio & Prediction$^1$\\
\hline 
A4     & 1.2172   & $>1.89$            & $N_{\rm H}<5$ [2.0]\\
       &          &                    & $N_{\rm H}<0.5$ [1.4]\\
A8     & 0.8030   & $>4.71^\ddag$      & $N_{\rm H}<0.1$ [2.0]  \ and \   $\Gamma>1.4$\\
A10    & 0.2242   & $>31.26$           & $\Gamma>3$\\
A12$^2$& 2.28     & $2.03\pm 1.1$      & $1<N_{\rm H}<30$ [2.0]\\
       &          &                    & $N_{\rm H}<20$ [1.4]  \ or$^*$\   $\Gamma>1.4$\\
A12$^2$& 0.66     & $2.03\pm 1.1$      & $0.2<N_{\rm H}<4$ [2.0]\\
       &          &                    & $N_{\rm H}<2$ [1.4]   \ or$^*$\   $\Gamma>1.4$\\
A13    & 1.5933   & $>5.9$             & $\Gamma>2$\\
A14    & 1.35     & $>2.20$	       & $N_{\rm H}<5$ [2.0]  \ and \   $\Gamma>1.4$\\
A17    & 1.466    & $>4.42^\ddag$      & $N_{\rm H}<0.5$ [2.0]  \ and \   $\Gamma>1.4$\\
A27    & 0.87     & $2.23\pm 1.0^\dag$ & $0.7<N_{\rm H}<4$ [2.0]\\
       &          &                    & $N_{\rm H}<2$ [1.4]   \ or$^*$\   $\Gamma>1.4$\\
       &          & $6.45\pm 2.6^\ddag$& $N_{\rm H}<0.4$ [2.0]   \ or$^*$\   $\Gamma>2.0$\\
A28    & 1.0330   & $4.38\pm 2.1^\dag$ & $N_{\rm H}<2$ [2.0]   \ or$^*$\   $\Gamma>2.0$\\
       &          &                    & $N_{\rm H}<0.2$ [1.4]   \ or$^*$\   $\Gamma>1.4$\\
\hline
\end{tabular}
~~~~~~\\
Redshifts stated to four decimal places are spectroscopic and the rest are photometric.\\
$^1$ The limits on the column densities (stated in units of $10^{22}$ cm$^{-2}$) account for the errors on the S/H ratio where available; otherwise have been constrained to the nearest half-decade. The assumed powerlaw-indices($\Gamma$) are stated in square brackets.\\
$^2$ The two rows correspond to the two possible photometric redshifts for this source. See Table~\ref{tab:zphot} and the text.\\
$^\dag$ Obtained from 800008 observation only.\\
$^\ddag$ Obtained from 800009 observation only.\\
$^*$ These two possibilities come from the values of S/H at the extremes of the range encompassed by the S/H-error. The upper-limit on $N_{\rm H}$ corresponds to the lower value of S/H and the lower-limit on $\Gamma$ corresponds to the higher value of S/H.\\
\end{table}


\twocolumn

\twocolumn



\begin{thebibliography}{}
\bibitem [] {} Allen D.A., 1976, MNRAS, 174, 29P
\bibitem [] {} Allen S.W., Ettori S., Fabian A.C., 2001, MNRAS
accepted (astro/ph-0008517)
\bibitem [] {} Altieri B., Metcalfe L., Kneib J.P., McBreen B. et al., 1999, A\&A, 343, L65
\bibitem [] {} Arnaud K.A., 1996, "Astronomical Data Analysis Software and
Systems V", eds. Jacoby G. and Barnes J., ASP Conf. Series vol. 101, 17
\bibitem [] {} Barger A.J, Cowie L.L., Mushotzky R.F., Richards E.A.,
2001, AJ, 121, 662
\bibitem [] {} Bolzonella M., Miralles J.-M., Pello R.,  2000, A\& A, 
363, 476.
\bibitem [] {} Brandt W.N. et al 2001 (astro-ph/0102411)
\bibitem [] {} Bruzual G.A, Charlot S., 1993, ApJ, 405, 538
\bibitem [] {} Calzetti D., Armus L., Bohlin R.C., Kinney A.L.,
Koornneef J., Storchi-Bergmann T., 2000, ApJ, 533, 682
\bibitem [] {} Coleman G.D., Wu C.-C., Weedman D.W., 1980, ApJS, 43, 393 
\bibitem [] {} Comastri A., Setti G., Zamorani G., Hasinger G., 1995, A\&A, 296, 1 
\bibitem [] {} Cowie L.L et al 2001, ApJ, 551, L9 
\bibitem [] {} Crawford C.S., Fabian A.C., Gandhi P., Wilman R.J.,
Johnstone R.M., 2001, MNRAS in press 
\bibitem [] {} Eales S.A., Rawling S., Dickinson M., Spinrad H., Hill G.J., Lacy M.,
         1993, ApJ, 409, 578
\bibitem [] {} Epps H.W., Miller J.S., 1998, Proc SPIE, 3355,48 
\bibitem [] {} Fabian A.C. et al 2000,  MNRAS, 315, L8
\bibitem [] {} Francis P.J., Hewett P.C., Foltz C.B., Chaffee F.H.,
Weymann R.J., Morris S.L., 1991, ApJ, 373, 465 
\bibitem [] {} Giacconi R. et al 2001 (astro-ph/0007240)
\bibitem [] {} Granato G.L., Danese L., Franceschini A., 1997, ApJ,
485, 147 
\bibitem [] {} Hasinger G. et al 1998, A\&A, 329, 482
\bibitem [] {} Hornschemeier A.E. et al 2001 (astro-ph/0101494)
\bibitem [] {} Ivezi\'{c} Z., Nenkova M., Elitzur M., 1999 (astro-ph/9910475)
\bibitem []{} Lehmann I. et al, 2001 (astro-ph/0103368)
\bibitem [] {} Lemonon L., Pierre M., Cesarsky C.J., Elbaz D., Pello
R., Soucail G., Vigroux L., 1998, A\&A, 334, L21
\bibitem [] {} Lilly S.J., Tresse L., Hammer F., Crampton D., Le Fevre,
O., 1995, ApJ, 455, 108
\bibitem [] {} Madau P., Ghisellini G., Fabian A.C., 1994, MNRAS, 270, L17
\bibitem [] {} Maiolino R. Marconi A., Salvati M., Risaliti G.,
Severgnini P., Oliva E., La Franca F., Vanzi L., 2001, A\&A, 365, 28
\bibitem [] {} Mushotzky R.F., Cowie L.L., Barger A. Arnaud K.A., 2000, Nature, 404, 459
\bibitem [] {} Setti G.,  Woltjer L., 1989, A\&A, 224, L21 
\bibitem [] {} Stark A.A., Gammie C.F., Wilson R.W., Bally J., Linke R.A.,
              Heiles C., Hurwitz M., 1992, ApJS, 79, 77
\bibitem [] {} Tozzi P. et al 2001 (astro-ph/0103014)
\bibitem [] {} Willott C.J. et al, 2001 (astro-ph/0105560) 
\bibitem [] {} Wilman R.J., Fabian A.C., 1999, MNRAS, 309, 862
\bibitem [] {} Wilman R.J., Fabian A.C., Gandhi P., 2000, MNRAS, 318, L11
\bibitem [] {} Wilman R.J., Fabian A.C., Nulsen P.E.J., 2000, MNRAS, 319,583
\bibitem [] {} Yee H.K.C. et al 1991, A\&AS, 88, 133
\end{thebibliography}
\end{document}